\documentclass[10pt, onecolumn]{extarticle}

\usepackage{xcolor}

\usepackage{soul}

\usepackage[bordercolor=white,backgroundcolor=gray!30,linecolor=black,colorinlistoftodos, textsize=10pt]{todonotes}

\usepackage{amssymb}
\usepackage{mathtools}
\usepackage{textcomp}
\usepackage{upgreek}
\usepackage{amsmath}
\usepackage[textstyle,amssymb]{SIunits}
\usepackage{graphicx}
\usepackage[normalem]{ulem}  
\usepackage{soul}
\usepackage[twocolumn,textwidth=183mm,columnsep=6mm,top=20mm,bottom=25mm]{geometry}
\usepackage{helvet}
\usepackage{titlesec}
\titleformat*{\section}{\bfseries\sffamily}
\titlespacing{\section}{0pt}{*4}{*0}
\titleformat{\subsection}[runin]{\normalfont\bfseries}{\thesubsection.}{3pt}{}
\usepackage{setspace}
\usepackage[breaklinks=true]{hyperref} 
\usepackage[hang,flushmargin]{footmisc} 
\usepackage[font={footnotesize,sf},labelfont={sf,bf},labelsep=endash,justification=raggedright]{caption}

\usepackage{natbib}
\usepackage[labelfont=bf,justification=justified]{caption}
\begin{document}

\twocolumn[\begin{@twocolumnfalse}
	{\fontsize{22pt}{22pt} \sf  \bfseries Sub-nominal resolution Fourier transform \\spectrometry with chip-based combs \par}
	\vspace{0.5cm}
	
	{\sf\large \textbf {Lukasz A. Sterczewski$^{1,2,*}$, and Mahmood~Bagheri$^{1,\mathparagraph}$}}
		\vspace{0.5cm}
		
		{\sf \textbf{$^1$Jet Propulsion Laboratory, California Institute of Technology, Pasadena, CA 91109, USA\\
		$^2$Faculty of Electronics, Photonics and Microsystems, Wroclaw University of Science and Technology, Wyb. Wyspianskiego 27, 50-370 Wroclaw, Poland\\
		\small$^*$e-mail: {lukasz.sterczewski@pwr.edu.pl}\\
		$^\mathparagraph$e-mail: {mahmood.bagheri@jpl.nasa.gov}
		}}
		\vspace{0.5cm}
\end{@twocolumnfalse}]
\vspace{0.5cm}

{\noindent \sf \small \textbf{\boldmath Chip-based optical frequency combs address the demand for compact, bright, coherent light sources of equidistant phase-locked lines. Traditionally, the Fourier Transform Spectroscopy (FTS) technique has been considered a suboptimal choice for resolving comb lines in chip-based sensing applications due to the requirement of long optical delays, and spectral distortion from the instrumental line shape. Here, we develop a sub-nominal resolution FTS technique that precisely extracts the comb's offset frequency in any spectral region directly from the measured interferogram without resorting to nonlinear $f$-to-$2f$ interferometry. This in turn enables MHz-resolution spectrometry with millimeter optical retardations. Low-pressure MHz-wide absorption lines probed by widely-tunable chip-scale mid-infrared OFCs with electrical pumping are fully resolved over a span of tens of nanometers. This versatile technique paves the way for compact, electrostatically-actuated, or even all-on-chip high-fidelity FTS, and can be readily applied to boost the resolution of existing commercial instruments several hundred times.}
}

The compact footprint, low power consumption and native operation in spectral regions relevant for optical sensing makes chip-scale optical frequency combs (OFCs)~\cite{scalari_-chip_2019, kippenbergMicroresonatorBasedOpticalFrequency2011, hugiMidinfraredFrequencyComb2012, bagheriPassivelyModelockedInterband2018a} ideal candidates for broadband and high-resolution spectrometers~\cite{yuMicroresonatorbasedHighresolutionGas2017}. Arguably, the most popular technique to resolve comb lines relies on dual-comb beating between a pair of mismatched combs~\cite{coddington_dual-comb_2016} on a fast microwave-bandwidth photodetector. Although dual-comb spectroscopy (DCS) enables real-time monitoring of broad optical bandwidths, it poses a significant challenge for precise, extended-timescale measurements. The difficulty lies in ensuring high mutual coherence between the sources via analog~\cite{coddington_coherent_2008,villares_dual-comb_2014,westberg_mid-infrared_2017} or digital synchronization schemes~\cite{burghoff_computational_2016,hebert_self-corrected_2017,sterczewskiComputationalCoherentAveraging2019}, and the need for high-speed digital signal processing. Additionally, strict requirements on the comb optical linewidth make many devices incompatible with the DCS technique. Another challenge faced by chip-scale OFCs is gap-less tunability~\cite{villares_dual-comb_2014, sterczewskiMultiheterodyneSpectroscopyUsing2017, gianella2020high} for performing measurements beyond the coarse mode spacing on the order of 5--20~GHz, which requires the presence of spectrally-matched low-phase noise regimes. 

Here, we show that a solution to precise high-resolution measurements using chip-scale OFCs lies in the Fourier Transform Spectroscopy (FTS) technique~\cite{griffiths2007fourier}, which fundamentally requires an identical integration time as the DCS technique to reach the same signal-to-noise ratio (SNR)~\cite{newbury2010sensitivity}, yet needs only a single comb. The high brightness of OFCs~\cite{udem2002optical, picqueFrequencyCombSpectroscopy2019} previously explored in conventional FTS systems ~\cite{mandon2007femtosecond,mandon2009fourier} is now merged with MHz resolutions obtainable at millimeters of optical retardations, which exceeds 100$\times$ the nominal resolution. All that is possible to implement in an arbitrary spectral region even for free-running OFCs. The only requirement is the knowledge of the comb's repetition rate, which can be obtained directly from the device's electrical bias or a photodetector.

\begin{figure}[!t]
	\centering
	\includegraphics[width=0.5\textwidth]{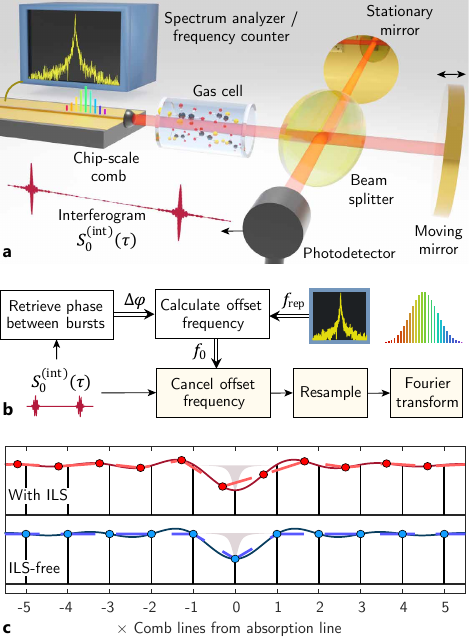}
	\caption{\textbf{Principle of chip-based self-enabled subnominal FTS}. \textbf{a,} Experimental setup constituting a Michelson interferometer with millimeter-long displacements of the moving mirror. An analogous beam path is for reference wavelength interferometry to ensure uniform sampling of the IGM. The spectrum analyzer / frequency counter can be removed once the repetition rate is characterized. \textbf{b,} Flowchart of the interferometric signal to meet the sub-nominal resolution criterion. \textbf{c,} Illustration of the sampling effect on the measured spectrum (single-burst interferogram, inspired by   Ref.~\cite{maslowskiSurpassingPathlimitedResolution2016}) when the spectrometer resolution is higher than $f_\mathrm{r}$ and $f_0$ is not removed (with ILS), and when points sampled by FTS are aligned with comb teeth positions (ILS-free / subnominal). Vertical lines represent comb lines, while filled dots correspond to sampled points. Although only the center comb line is absorbed (Lorentz profile), the ringing artefact affects multiple lines, which disappears only upon matching the FTS sampling and comb frequency scales. 
	\label{fig:Schematic} }
\end{figure}

Whereas the apparent modal sparsity of chip-scale OFCs can be seen as a disadvantage, it becomes a key enabler for compact high-resolution FTS. The FTS technique employs a Michelson or Mach-Zehnder interferometer to measure a time-averaged field autocorrelation known as the interferogram (IGM, $S_0^{(\mathrm{int})}$) related to the power spectral density via the Fourier transform. The IGM results from optical interference between the optical source's waveform and its time-delayed copy on a slow photodetector. A typical IGM of an OFC is expected to possess intensive bursts separated by lower-intensity regions (Fig.~\ref{fig:Schematic}a) with envelope periodicity relating to the inverse of the comb repetition rate $f_\mathrm{r}$.

For an arbitrary (even incoherent) source, the nominal spectral resolution $\Delta \widetilde{\nu}_\mathrm{min}$ in FTS is limited by the inverse of the maximum optical path difference (OPD) $\Delta_\mathrm{max}$ between the interferometer moving and stationary arms. However, the delay range resolution limit can be greatly surpassed if the light source has OFC properties~\cite{maslowskiSurpassingPathlimitedResolution2016, rutkowskiOpticalFrequencyComb2018}. A special IGM sampling procedure can virtually eliminate a convolution of the measured spectrum with an instrumental line shape (ILS) function induced by a truncation of the scanned optical path. The truncation acts as a box-car window, which limits the resolution and distorts the measured spectrum. This can be mitigated via ILS-free FTS: one has to measure exactly one period of the IGM, which implies an OPD of $\Delta_\mathrm{max}=c/f_\mathrm{r}$. This is why a 10~GHz semiconductor laser OFC requires only $\sim 3$~cm of OPD for megahertz resolution FTS spectra, which translates into 15~mm of the moving arm displacement in conventional, and 7.5~mm in double-sided mirror~\cite{yangPrincipleAnalysisMoving2012} or double-pass~\cite{jenningsDoublePassingKitt1985} configurations, or even sub-mm in double-pendulum arrangements~\cite{jaacksDoublePendulumMichelson1989}. This is in stark contrast to nominal MHz-resolution FTS, which requires an OPD in the range of meters.

Unfortunately, the indispensable requirement of prior ILS-free FTS approaches is the knowledge (and stabilization) of the comb's carrier-envelope-offset (CEO), or simply offset frequency $f_0$, which traditionally has been obtained via $f$-to-$2f$ interferometry~\cite{telle1999carrier} of an octave-spanning source. Since in many spectral regions the $f$-to-$2f$ scheme is either impractical or virtually impossible to implement, this limitation has excluded spectrally narrower sources with low pulse energies or with frequency-modulated (FM) emission spectra from surpassing the nominal FTS resolution, which constitute a majority of chip-scale OFCs. From a system complexity standpoint, bulky stabilized near-infrared fiber OFCs with MHz repetition rates require meters of OPD to meet the subnominal resolution criterion, which has restricted such instruments to a laboratory environment. Additionally, access to the spectroscopically-relevant mid-infrared spectral range has relied on nonlinear frequency conversion techniques~\cite{khodabakhshFourierTransformVernier2016}, which further increased the system complexity and footprint. 
By adapting the subnominal FTS procedure to electrically-tunable chip-scale OFCs without an easily measurable offset frequency, we perform high-resolution (MHz) and broadband (THz-wide) spectroscopy with millimeter-long mechanical displacements using millimeter-scale free-running sources operating natively in the mid-infrared. The technique may find application in existing Fourier Transform Infrared Spectrometers (FTIRs), which have conventionally been seen as incompatible with high-resolution applications. Our technique (inspired by prior work of Maslowski et al. in Ref.~\cite{maslowskiSurpassingPathlimitedResolution2016}, where all comb parameters needed to be known and stabilized) enables turning them into instruments with equivalent OPDs on the order of meters without any modifications. This in turn gives access to Doppler-limited molecular transitions at lower pressures and temperatures such as occurring in space. Also, recent developments in on-chip FTS with millimeter-scale displacements~\cite{zhang2021research} can be leveraged to perform high-resolution on-chip spectrometry with meter-scale equivalent OPDs.

\section*{Experiment}
\noindent We will now demonstrate the application of the sub-nominal FTS technique to interleaved measurements of low-pressure gaseous analytes performed using a compact Michelson interferometer (Fig.~\ref{fig:Schematic}a). Here, we use two different mid-infrared chip-scale OFCs: a widely-tunable interband cascade laser (ICL) comb, and a quantum well diode laser (QWDL) comb (see Supplementary Information for details). A full mathematical description of the technique is given in Methods, but the general idea is laid out here. The essence of the proposed ILS-free FTS procedure (Fig.~\ref{fig:Schematic}b) is to modify the measured single-period IGM so that comb lines in the frequency domain are located at zero-crossings of the truncation-induced ILS~\cite{maslowskiSurpassingPathlimitedResolution2016, rutkowskiOpticalFrequencyComb2018}. This is shown schematically in Fig.~\ref{fig:Schematic}c, which clearly shows what happens when this condition if fulfilled (ILS-free) and violated (with ILS). Conventionally, a prerequisite for implementing this technique is the knowledge of the comb's $f_\mathrm{r}$ and $f_0$. The ILS-cancelling routine digitally nulls $f_0$ followed by IGM resampling and truncation to include exactly one signal period defined by $f_\mathrm{r}$. Unfortunately, whereas $f_\mathrm{r}$ is easily measurable directly from a laser cavity or photodetector (with $10^{-6}$ or higher precision), for most electrically-pumped OFCs it is extremely challenging to measure $f_\mathrm{0}$. Fortunately, the combination of $f_\mathrm{r}$ knowledge with digital retrieval of the IGM phase increment allows us to bypass this limitation. The key is to acquire an asymmetric \emph{single-sided} IGM that carries information about $f_0$ instead of the double-sided used conventionally. This technique unlocks the high-resolution FTS potential of many comb platforms, where $f_\mathrm{0}$ retrieval via $f$-to-$2f$ interferometry~\cite{telle1999carrier} would be difficult or virtually impossible to implement. By definition, the IGM centerburst has zero phase at zero delay (constructive interference of all wavelengths). However, the peak of the first IGM satellite (at a temporal delay of $1/f_\mathrm{r}$) accumulates a phase related to $f_0/f_\mathrm{r}$. When corrected for the discrete number of points in the IGM (which are sampled at multiples of the FTS reference wavelength $\lambda_\mathrm{ref}$), it can be used as a MHz-accurate estimate of $f_\mathrm{0}$.

\begin{figure}[!tb]
	\centering
	\includegraphics[width=0.5\textwidth]{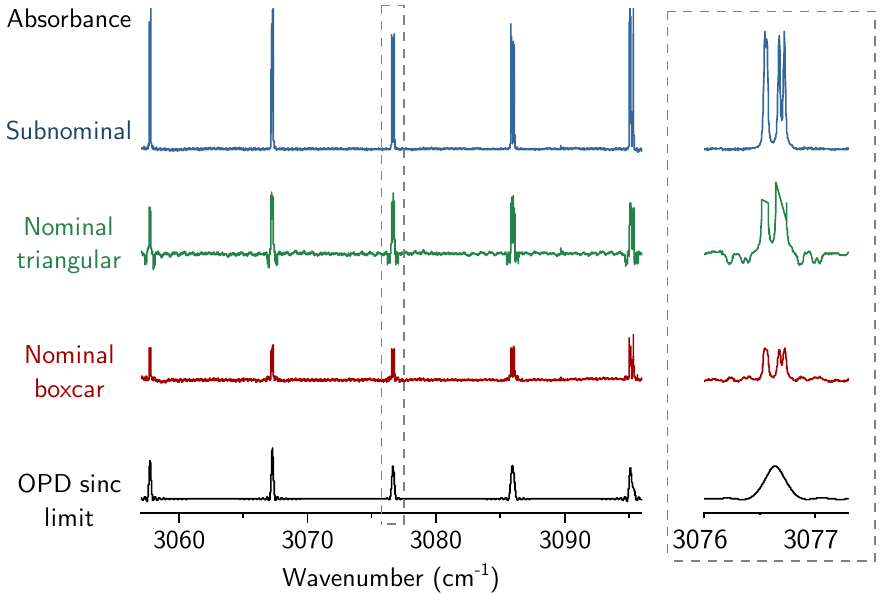}
	\caption{\textbf{Comparison of conventional FTS measurements with the subnominal technique.} The measured analyte is methane (CH$_4$) in natural isotopic ratio at 95~Torr (12.66~kPa) and room temperature, which displays complex manifold transitions with MHz linewidths at such conditions. In the nominal resolution case, capturing 2 bursts yields a mode-resolved comb spectrum with peak intensities used for spectroscopy. Although in the unapodized (box-car-windowed) and triangularly-apodized case the resolution is higher than for an incoherent source (OPD sinc limit), the lines are severely distorted. Ringing artifacts, negative absorbance, and peak rounding appear. In contrast, the sub-nominal FTS technique provides an undistorted, high-fidelity spectrum comparable with an instrument with a 100$\times$ higher resolution.}
	\label{fig:Comparison}
\end{figure}

\begin{figure*}[!tb]
	\centering
	\includegraphics[width=1\textwidth]{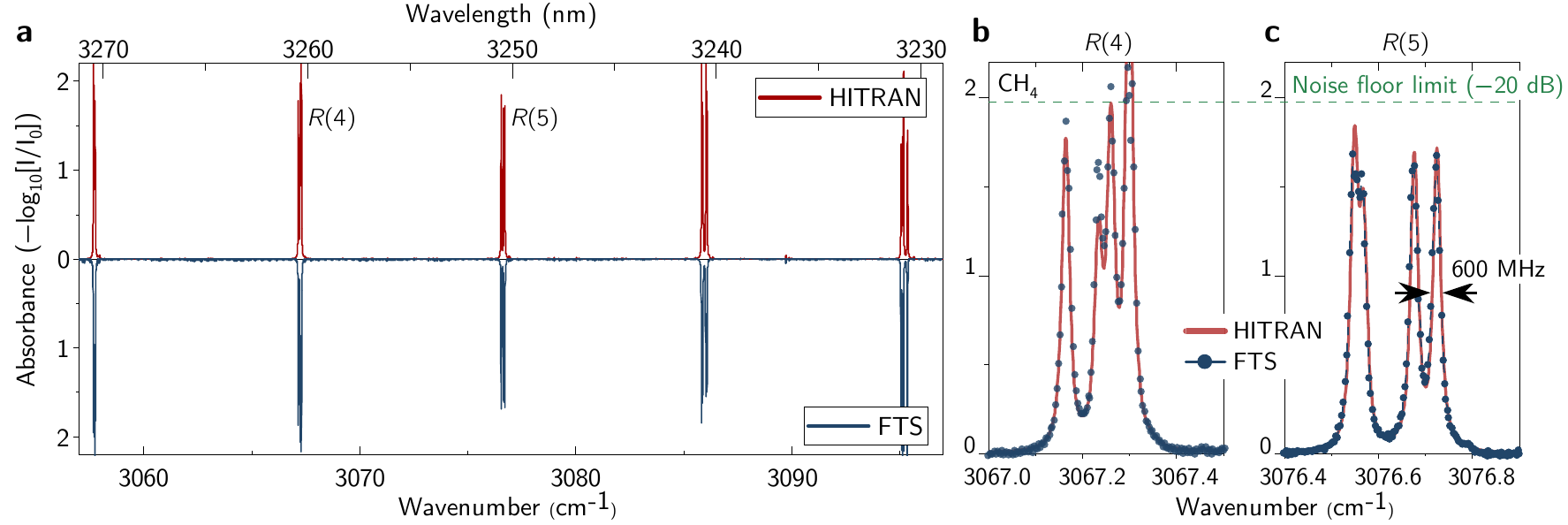}
	\caption{\textbf{The $\nu_3$ mid-infrared band of methane (CH$_4$) with 5 manifold $R$-branch transitions probed by subnominal FTS based on a 4-mm long comb tuned in injection current in steps of 96~MHz. The nominal FTS resolution is 9.6~GHz.} \textbf{a,} HITRAN 2020 fit along with the FTS measurement, \textbf{b,} Zoom onto line $R(4)$. At higher absorptions ($\geq$15~dB attenuation), the lowered SNR induces pronounced intensity fluctuations at absorption peaks, \textbf{c,} Zoom onto $R(5)$ line with 600-MHz-wide features.}
	\label{fig:methane}
\end{figure*}

\begin{figure*}[!tb]
	\centering
	\includegraphics[width=1\textwidth]{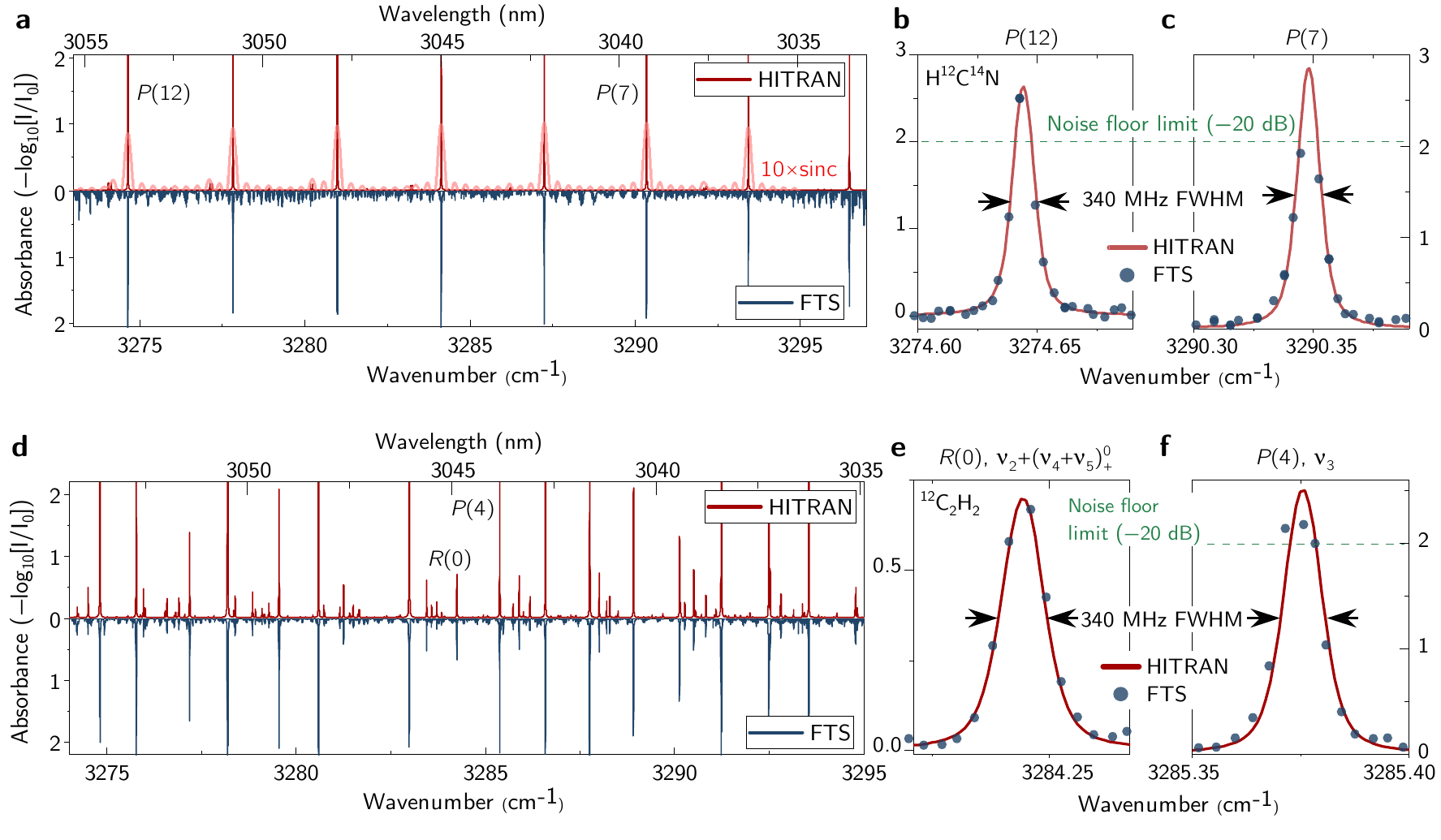}
	\caption{\textbf{Subnominal FTS based on a 4-mm long comb tuned in injection current in steps of 100~MHz. The nominal FTS resolution is 10~GHz.} \textbf{a,} 
	The $\nu_3$ mid-infrared band of hydrogen cyanide (H$^{12}$C$^{14}$N) with $P$-branch transitions probed by the system. The analyte pressure was 2~Torr. The subnominal FTS spectrum agrees well with a HITRAN 2020 database model. Additionally shown is a 10$\times$ magnified, simulated absorbance spectrum probed by a Michelson interferometer with a 3~cm OPD (sinc). Pronounced amounts of noise (0.1 in absorbance units) are due to a greater sensitivity of QWDL devices to to optical feedback. \textbf{b,} Zoom onto line $P(12)$ with a 340~MHz wide line. \textbf{c,} Zoom onto line $P(7)$. \textbf{d,} The fundamental $\nu_3$, and $\nu_2+(\nu_4+\nu_5)$ combination bands of acetylene ($^{12}$C$_2$H$_2$) at 10~Torr probed by the system compared with HITRAN. \textbf{e,} Zoom of line $R(0)$ in the combination band. \textbf{f,} Zoom of line $P(4)$ in the fundamental $\nu_3$ band. The dynamic range limitation (20~dB) clips the absorption.}
	\label{fig:HCN_C2H2}
\end{figure*}

\section*{Spectroscopy}
\noindent To prove the exactitude of the technique, we have compared the ILS-free spectra of 95~Torr methane (CH$_4$) with those obtained by conventional treatment of IGMs acquired by an interferometer with an OPD of $\sim$35 mm (8.5~GHz resolution), which captures the full centerburst, and one roundtrip burst. As an OFC source, we used an ICL comb~\cite{bagheriPassivelyModelockedInterband2018a} with a 9.6~GHz repetition rate tuned by the injection current in a gap-less fashion (over a full $f_\mathrm{r}$). Peak intensities and positions were tracked for line intensity data in conventional FTS for comparison. Fig.~\ref{fig:Comparison} shows the measured absorbance in the case of unapodized (nominal boxcar), apodized (triangular window), and proposed subnominal FTS technique. While the peak tracking approach enables us to bypass the conventional incoherent source resolution limit (OPD sinc limit), the lines remain nonetheless severely distorted. They possess ringing artifacts, and display negative absorbances, while some features remain unresolved with drastically reduced intensities. In contrast, the ILS-free data that required only the knowledge of the comb's $f_\mathrm{r}$ possess clear, well-resolved, and undistorted lines. A comparison of the full-span methane absorbance measurement including $>10^{4}$ spectral elements spaced by 96~MHz with the HITRAN 2020 database~\cite{gordon2022hitran2020} is shown in Fig.~\ref{fig:methane}a, and reveals excellent agreement between the fitted, and measured spectrum covering $>40~$cm$^{-1}$ (1.2~THz / 42~nm) with a 96~MHz point spacing. Manifold features of lines $R(4)$, and $R(5)$ in the $\nu_3$ band of CH$_4$ as narrow as 600~MHz at these conditions, are faithfully reproduced (Fig.~\ref{fig:methane}b, and Fig.~\ref{fig:methane}c, respectively). The major limitation results from the limited dynamic range of 20~dB (absorbance of 2), when the measurement noise floor is reached. Consequently, the SNR of comb lines probing such strong absorptions is drastically lower. It can be greatly improved with longer averaging (here it took 1.5 seconds per step), or a comb source with a greater power per mode.

To demonstrate the versatility of the technique and its independence on the comb source, we employed a recently demonstrated mid-IR quantum well diode laser (QWDL) comb~\cite{sterczewski2022battery} at 3~$\upmu$m to probe two pure molecular standards: hydrogen cyanide (H$^{12}$C$^{14}$N), and acetylene ($^{12}$C$_2$H$_2$), at pressures of 2~Torr, and 10~Torr, respectively. Both analytes possess $\sim$340~MHz wide Doppler-limited absorption lines (FWHM), which practically require an instrumental resolution greater than 100~MHz to accurately reproduce the lineshape. Figure~\ref{fig:HCN_C2H2}a plots the subnominal resolution FTS measurement of the fundamental $\nu_3$ mid-infrared band of pure H$^{12}$C$^{14}$N, which displays pronounced amounts of noise (0.1 in absorbance units) due to a greater sensitivity to optical feedback of QWDL devices (see Methods). Also, the QWDL comb has a narrower span ($\sim$20~nm). Nevertheless, the FTS measurement is of high fidelity, as shown in Fig.~\ref{fig:HCN_C2H2}b for line $P(12)$, and Fig.~\ref{fig:HCN_C2H2}c for $P(7)$, except for weak absorbance clipping to a value of 2 due to the SNR limit. Even more complex features due to overlapping spectroscopic bands are well resolved for the second molecular standard of $^{12}$C$_2$H$_2$ at 2~Torr (Fig.~\ref{fig:HCN_C2H2}d) with two representative lines in the low-absorbance (Fig.~\ref{fig:HCN_C2H2}d), and high-absorbance regime (Fig.~\ref{fig:HCN_C2H2}e), both in excellent agreement with the HITRAN model. Note that in all measurements, the spectroscopic axis was retrieved solely from the measured repetition rate and a known interferometer reference wavelength (here HeNe laser). The uncertainty of the frequency scale retrieved from consecutive measurements is estimated to be $\pm$0.005~cm$^{-1}$ (150~MHz, 1.5\% relative to $f_\mathrm{r}$). Note that the feasibly of exactly matching the spectrometer sampling points to the comb peak locations is in stark contrast to comb spectroscopy based on grating optical spectrum analyzers (OSA)~\cite{yuMicroresonatorbasedHighresolutionGas2017}, which additionally lacks the Jacquinot- (throughput), Fellgett- (multiplex), and
Connes- (wavelength accuracy) advantages of FTS. To date, high-resolution OSA instruments have been used to probe only multi-GHz absorption lines, while distortion effects arising due to this mismatch and entrance slit effects may become dominant in the Doppler-limited regime, as probed here. 

Although in this proof-of-concept, minute-scale demonstration we utilize a commercial FTIR instrument, the combination of millimeter-long optical displacements with the electrically-pumped chip-scale source paves the way for extremely compact, precise, battery-operated spectroscopic instruments with chip-size footprints~\cite{zhang2021research}. Even existing on-chip spectrometers may be boosted in resolution almost a thousand times by employing widely-tunable shorter-cavity OFCs with repetitions on the order of tens of GHz like microresonators~\cite{kippenbergMicroresonatorBasedOpticalFrequency2011} or quantum cascade lasers (QCLs)~\cite{scalari_-chip_2019, hugiMidinfraredFrequencyComb2012}. Obviously, the sub-nominal technique does not change their coarse, GHz spectral sampling grid. It only ensures that sparse, interleaved spectra with MHz resolutions around the comb teeth are undistorted. For electrically-pumped OFC, the need for comb line positions tuning for spectral interleaving can be easily fulfilled by simply changing the injection current or temperature. This motivates the future development of broadband OFC sources with reproducible gap-less tuning capabilities yet with less focus on absolute frequency stability. This is because the proposed technique unlocks the high-resolution spectroscopy potential of emerging OFC chips with DCS-incompatible optical linewidths. For instance, in some regimes, the ICL comb used here exhibits a comb linewidth of $\sim$50~MHz, which would render an unresolved and noisy microwave spectrum with the DCS technique, but performs well in FTS. Analogous challenges are faced by QCLs operating in the terahertz range~\cite{burghoff2014terahertz}. We also envision enhancing the digitally-enabled ILS-free technique by laser modulation schemes to sense low concentrations of analytes without relying on direct absorption measurements. Beyond analytical spectroscopy, the technique may also find application in OFC coherence characterization using linear microwave interferometry techniques such as SWIFTS~\cite{burghoff2015evaluating} to completely eliminate the need for lineshape deconvolution. It will also enable one to analyze the offset frequency tuning characteristics and dynamics in emerging OFC platforms without resorting to optical heterodyne techniques, which are difficult to implement in more exotic spectral regions. 

\section*{Methods}

\vspace{0.2cm}
\footnotesize
\setstretch{1.}
\noindent\textbf{Offset frequency retrieval}: 
The electric field emitted by an optical frequency comb can be expressed as a superposition of its equidistant lines
\begin{equation}
E(t) = \sum_{n} E_n \mathrm{e}^{\mathrm{i}(\omega_n t + \varphi_n)}\, ,
\label{eq:FTS}
\end{equation}
where $E_n$ is the $n$-th comb line intensity, $\omega_n=2\mathrm{\pi}f_n$ is the angular optical frequency, and $\varphi_n$ is the phase. $\omega_n$ obviously follows the frequency comb model with a common offset frequency $\omega_0$, and a repetition frequency $\omega_\mathrm{r}$ such that $\omega_n=\omega_0+n\omega_\mathrm{r}$. The summation in Eq.~\ref{eq:FTS} is over both positive and negative $n$ because the field is real, which implies that $E_n$=$E^*_{-n}$ and $\omega_n=-\omega_{-n}$.

In FTS, the measured interferometric quantity is the autocorrelation of the time-averaged electric field $E(t)$, which disregards the optical phase:

\begin{equation}
S_0^{(\mathrm{int})}(\tau) = \sum_{n} \langle |E_n|^2 \rangle \mathrm{e}^{\mathrm{i}\omega_n \tau} \, .
\label{eq:FTS2}
\end{equation}
The autocorrelation $S_0^{(\mathrm{int})}(\tau)$, referred to as the IGM, is a function of a relative optical delay $\tau=\Delta/c$. For an infinitely long acquisition using a perfect Michelson interferometer, the Fourier transform of Eq.~\ref{eq:FTS2} yields an array of Dirac deltas located at $\omega_n$ with intensities $\langle |E_n|^2 \rangle$. Real interferometers, however, have limited optical displacements ($|\Delta|<\infty$), which introduce a truncation of the acquisition, corresponding in the frequency domain to a convolution of the true optical spectrum with an ILS function. For short optical displacements, not only will it limit the spectral resolution, but also introduce ringing artifacts from absorption lines. Fortunately, the influence of the ILS can be suppressed if one uses an optical frequency comb source~\cite{maslowskiSurpassingPathlimitedResolution2016}. Note that in contrast to earlier approaches that correct the IGM based on known and stabilized $f_\mathrm{r}=\omega_\mathrm{r}/2\mathrm{\pi}$ and $f_0=\omega_0/2\mathrm{\pi}$ to obtain an ILS-free spectrum, here we are solving a quasi-inverse problem. Given a measurement of free-running $f_\mathrm{r}$, we \emph{estimate} $f_\mathrm{0}$ directly from the IGM, which is next digitally removed. Both frequencies are then used for frequency axis calibration.  

At the heart of the ILS-free technique proposed here is the idea to convert a single-sided interferogram measured over $\tau\in [0,T_\mathrm{r}]$, where $T_\mathrm{r}=1/f_\mathrm{r}$, into a harmonic signal so that it can be circularly shifted just like it was recorded in double-sided mode ($\tau\in [-T_\mathrm{r}/2,T_\mathrm{r}/2]$). In fact, the measured interferogram $S_0^{(\mathrm{int})}(\tau)$ can be seen as a carrier wave $\omega_c$ modulated by a periodic envelope function. While the phase of the centerburst (around $\tau=0$) is zero (by definition of the autocorrelation requiring to have a global maximum at $\tau=0$), satellite bursts possess a relative phase shift $\Delta \varphi$ between the envelope peak and the carrier, which is directly related to the offset frequency. The measured 1-st satellite (roundtrip) IGM burst can be expressed as:

\begin{equation}
S_0^{(\mathrm{int})}(\tau \pm T_\mathrm{r}) = \sum_{n} \langle |E_n|^2 \rangle \mathrm{e}^{\mathrm{i}\omega_n (\tau \pm T_\mathrm{r})} \, .
\label{eq:FTS3}
\end{equation}
Periodicity of the interferogram $S_0^{(\mathrm{int})}(\tau \pm T_\mathrm{r})=S_0^{(\mathrm{int})}(\tau)$ every $T_\mathrm{r}$ takes place only for harmonic (offset-free) combs i.e. such that have $\omega_0=0$ because $\mathrm{e}^{\mathrm{i}\omega_\mathrm{r}T_\mathrm{r}}=1$. Therefore, $\omega_0$ needs to be estimated and digitally removed from the IGM via frequency shifting.

To computationally extract the angular offset frequency $\omega_0$ much more precisely than via interpolating peaks in the frequency spectrum of a long interferometer scan covering multiple interferogram bursts, it is sufficient to retrieve the phase increment $\Delta \varphi$ between the complex interferogram centerburst and the first satellite occurring after $T_\mathrm{r}$, namely:
\begin{equation}
\Delta \varphi = \mathrm{Arg}\Bigg(\frac{\widetilde{S}_0^{(\mathrm{int})}(\tau+T_\mathrm{r})}{\widetilde{S}_0^{(\mathrm{int})}(\tau)}\Bigg) = \omega_0 T_\mathrm{r} + \underbrace{\omega_\mathrm{r}T_\mathrm{r}}_\text{$2\uppi$} \,.
\label{eq:FTS4}
\end{equation}
where the real IGM $S_0^{(\mathrm{int})}(\tau)$ has been converted into an analytic signal $\widetilde{S}_0^{(\mathrm{int})}(\tau)$ via the Hilbert transform $\mathcal{H}\{\ldots\}$
\begin{equation}
\widetilde{S}_0^{(\mathrm{int})}(\tau) = S_0^{(\mathrm{int})}(\tau) + \mathrm{i}\mathcal{H}\{S_0^{(\mathrm{int})}(\tau)\} \;.
\label{eq:FTS5}
\end{equation}
It is easy to realize that $\Delta \varphi_0=\Delta \varphi~\mathrm{mod}~2\uppi$ if one recalls the definition of the offset frequency $f_0=\frac{\Delta \varphi_0 ~\mathrm{mod}~2\uppi}{2\uppi}f_\mathrm{r}$,
which for the angular frequency is simply $\omega_0=\Delta \varphi_0 f_\mathrm{r}$. Therefore, $\omega_0 T_\mathrm{r}=\Delta \varphi_0$. 

The power of this $f_0$ retrieval technique stems from the fact that even for a free-running laser, $T_\mathrm{r}$ is typically known with relatively high precision ($10^{-6}$ or higher), which allows one to incorporate this knowledge to obtain $f_0$ directly from acquired single-period IGMs. The exactitude of the retrieval will predominantly depend on the accuracy of the reference laser frequency used for measuring the optical displacement, and that of $f_\mathrm{r}$. It should be also noted that this technique is restricted to probe only slow offset temporal dynamics, as only time-averaged signals are measured, yet it is still sufficient for the retrieval of correction parameters for near-second acquisition timescales of FTS. 

What follows from Eq.~\ref{eq:FTS4} is that in principle only \emph{one point} is sufficient to retrieve $\Delta \varphi$; however, for a statistically more accurate estimate, it is practical to calculate the mean of many samples. Rather than calculating the mean of complex arguments, which would be statistically biased due to the nonlinear operation of the arctangent function and issues with phase unwrapping, the correct way would be to average complex vectors $\tilde{S}_0^{(\mathrm{int})}(\tau+T_\mathrm{r})/\tilde{S}_0^{(\mathrm{int})}(\tau)$ first, and then calculate the complex argument of the average only. 

\vspace{0.2cm}
\noindent\textbf{Dealing with discrete number of points}: Because the actual interferometer displacement is constrained to be a multiple of the reference wavelength $\lambda_\mathrm{ref}$, the offset phase increment
$\Delta \varphi_0$ will almost never be measured at exactly $T_\mathrm{r}$. This also means that the natural frequency spacing in FTS $f_\mathrm{FTS}=c/(N\lambda_\mathrm{ref})$ will be close, but not perfectly matched to that of the comb $f_\mathrm{r}$, where $N$ is the number of acquired IGM points. Fortunately, it can be corrected to arbitrary precision by means of Fourier interpolation, i.~e. the IGM can have an arbitrary length $N'$ with an equivalent change of $\lambda_\mathrm{ref}'=N/N'$ also known as resampling. Simple IGM interpolation $q$ times also helps to suppress errors introduced by the discrete intervals of sampled data like the phase error resulting from missing the peak of the centerburst occurring at $\Delta=0$. Another advantage is that when the repetition rate varies throughout the scan, IGMs can be made of the same length (number of samples) through interpolation. This greatly simplifies the analysis of interleaved spectra.   

To ensure that after phase-shifting the IGM is symmetric (since the measured spectrum is real), one has to acquire $N=2N_\mathrm{ss}+1$ samples, $N_\mathrm{ss}$ unique on each side. This is equivalent to an IGM with one sample at ZPD ($\tau=0$) surrounded by $N_\mathrm{ss}$ samples on each side obtained through
\begin{equation}
N_\mathrm{ss} = \Bigg[\frac{f_\mathrm{r}}{2c\lambda_\mathrm{ref}} \Bigg]\;,
\label{eq:FTS_nPoints}
\end{equation}
where $[\ldots]$ stands for rounding to the nearest integer. A discrete Fourier transform (or more specifically, the Fast Fourier Transform / FFT) of the $N=2N_\mathrm{ss}+1$-samples-long signal will yield a frequency spectrum with $N_\mathrm{ss}+1$ points between DC and the Nyquist frequency $c/\lambda_\mathrm{r}$, or equivalently spaced by $f_\mathrm{FTS}$. Still, to avoid ILS distortion, the linear phase ramp in the IGM due to $f_0$ must be removed prior to FFT calculation.

To obtain an estimate of the offset frequency $\hat{f}_0$ given the discrete constraints, one has to retrieve the carrier phase increment for the $(k+N-1)$-th IGM point with that spaced $N-1$ points apart
\begin{equation}
\widehat{\Delta \varphi}[k] = \mathrm{Arg}\Bigg(\frac{\widetilde{S}_0^{(\mathrm{int})}[k+N-1]}{\widetilde{S}_0^{(\mathrm{int})}[k]}\Bigg) \,,
\label{eq:FTS_estimate1}
\end{equation}
assuming sample indexing starts at 1. The estimation exactitude can be greatly improved by averaging 
($\langle \ldots \rangle$) in the complex domain using IGM samples that have non-zero magnitudes:
\begin{equation}
\underset{\substack{k\,:\,|\widetilde{S}_0^{(\mathrm{int})}[k]| \gg 0\\ k\,:\,|\widetilde{S}_0^{(\mathrm{int})}[k+N-1]| \gg 0 }}{\langle \, \widehat{\Delta \varphi} \, \rangle} = \mathrm{Arg}\Bigg \langle \frac{\widetilde{S}_0^{(\mathrm{int})}[k+N-1]}{\widetilde{S}_0^{(\mathrm{int})}[k]}\Bigg \rangle  \,. 
\label{eq:FTS_estimate_av}
\end{equation}
Although this value is close to the true offset phase increment $\Delta \varphi_0$, the discrete domain constraints may introduce a significant estimation bias. The non-integer number of IGM points required for a bias-free estimate in the perfect ODL-comb match condition is:
\begin{equation}
N_{\not\mathbb{Z}}
\label{eq:N_points_required} = \frac{f_\mathrm{r}}{c\lambda_\mathrm{ref}} + 1 \;,
\end{equation}
while measured are $N$ discrete IGM points. Therefore, one has correct the $\langle \, \widehat{\Delta \varphi} \, \rangle$ estimate by a phase shift due to the $N_{\not\mathbb{Z}}-N$ mismatch dependent on the IGM carrier frequency. It can either be coarsely assumed to lie around the center of mass of the comb emission spectrum, or retrieved from the IGM. Here, we rely on the latter. Using the Kay frequency estimator that avoids phase unwrapping~\cite{kayFastAccurateSingle1989a}, $\omega_\mathrm{c}$ in units of cycles per samples is retrieved analogously to Eq.~\ref{eq:FTS_estimate_av}:
\begin{equation}
\underset{\substack{k\,:\,|\widetilde{S}_0^{(\mathrm{int})}[k]| \gg 0\\ k\,:\,|\widetilde{S}_0^{(\mathrm{int})}[k+1]| \gg 0 }}{\langle \, \widehat{\omega_\mathrm{c}} \, \rangle} = \mathrm{Arg}\Bigg \langle \frac{\widetilde{S}_0^{(\mathrm{int})}[k+1]}{\widetilde{S}_0^{(\mathrm{int})}[k]}\Bigg \rangle  \,. 
\label{eq:FTS_w_c_estimate_av}
\end{equation}
Oscillation at $\omega_\mathrm{c}$ over $N_{\not\mathbb{Z}}-N$ samples accumulates a phase
\begin{equation}
\delta \varphi = (N_{\not\mathbb{Z}}-N)\omega_\mathrm{c} \,.
\label{eq:FTS_discrete_correction_term}
\end{equation}
This correction term is included in the offset frequency estimate:
\begin{equation}
\widehat{f_0} = \frac{\langle \, \widehat{\Delta \varphi} \, \rangle + \delta \varphi}{2\uppi}f_\mathrm{r} \,.
\label{eq:FTS_estimate_final}
\end{equation}
Of course, the numerator must be phase-unwrapped to avoid large frequency jumps ($f_0$ typically evolves slowly, much slower than $f_\mathrm{r}$ per current step). Please also note that this frequency is used only for retrieving the frequency axis in the FTS measurement or studying the offset frequency evolution versus injection current. In the offset phase removal routine, however, it is better to multiply by a complex exponential with a linear phase ramp from 0 to $\widehat{\Delta \varphi}$ in the IGM domain, as discussed in the next section.

\vspace{0.2cm}
\footnotesize
\setstretch{1.}
\noindent\textbf{Preparation of the interferogram}: 
The offset frequency correction to ensure signal harmonicity relies on complex multiplication. The analytic signal is phase-shifted according to a phase ramp that increases linearly from 0 up to the $k/N$ scaled phase argument $\langle \, \widehat{\Delta \varphi} \, \rangle$ retrieved from the IGM as in Eq.~\ref{eq:FTS_estimate_av}
\begin{equation}
S_\mathrm{corr}[k] = \mathfrak{Re}\{\mathrm{e}^{-\mathrm{i} k/N\langle \, \widehat{\Delta \varphi} \, \rangle}\widetilde{S}_0^{(\mathrm{int})}(\tau)\}  \;,
\label{eq:FTS6}
\end{equation}
where $\mathfrak{Re}\{\ldots\}$ stands for the real part of a complex argument. Consequently, $\hat{f}_0$ is used only for retrieving the frequency axis rather than employed in the correction procedure. A visual depiction of the IGM at each processing step is provided in Supplementary Figure~S1.

\vspace{0.2cm}
\footnotesize
\setstretch{1.}
\noindent\textbf{Ambiguity of the offset frequency sign}: 
Because the IGM is a real signal, one cannot unambiguously determine the sign of $f_0$ solely from a single-period IGM. While it is not relevant for offset frequency cancellation, information if one deals with $+f_0$ or $-f_0$ is of critical importance for optical frequency axis calibration or for diagnostic purposes like studying the evolution of $f_0$ as a function of injection current or temperature. This problem is analogous to that of $f$-to-$2f$ interferometry, where between DC and $f_\mathrm{r}$ two microwave beat notes lie symmetrically between $f_\mathrm{r}/2$ (Nyquist frequency). Initially, one cannot distinguish between $+f_0$ and $-f_0$, until one actuates the laser to check in what directions they move. This piece of information enables one to determine if $f_0$ is positive or negative. 

This algorithm deals with the $\pm f_0$ ambiguity by incorporating prior knowledge about the tuning linearity of multimode (or comb) sources. Supplementary Figure~S3 shows that the incorrect sign of the retrieved $f_0$ magnifies the oscillatory frequency change due to $f_\mathrm{r}$ (which is measured electrically from the device). When guessed correctly, it suppresses frequency fluctuations and yields a smooth tuning curve. This behavior is typical for chip-scale emitters and has been validated (with coarse instrumental resolution), as shown in Supplementary Figures~S4 and S5.

\vspace{0.2cm}
\footnotesize
\setstretch{1.}
\noindent\textbf{Calculation of the frequency spectrum}: We calculate an $N$-point discrete Fourier transform (DFT) of $S_\mathrm{corr}$ using the FFT algorithm. This yields $N/2+1$ unique points starting from zero frequency (DC) spaced by in the optical domain $f_r$. The offset frequency is added globally for each comb line position, so that the $n$-th point of the ILS-free spectrum lies at an optical frequency
\begin{equation}
f_n^{\mathrm{FTS}}=nf_\mathrm{r}+\hat{f_0}  \;.
\label{eq:FrequencyAxis}
\end{equation}
Supplementary Figures S2 and S6 show the retrieved optical frequency for one individual comb line along with $f_0$ and $f_\mathrm{rep}$ as a function of injection current. Both comb platforms exhibit nearly linear tuning. 

The FFT is performed on an IGM that starts at $\Delta=0$ (ZPD) and lasts exactly one period. Because in some cases the centerburst is not sampled at exactly $\Delta=0$, there is a small phase error due to IGM's asymmetry, which implies phase correction. Such errors may also arise from residual offset phase persistent in the round trip burst. Here, a variation of the Mertz method is used~\cite{mertz1967auxiliary} to address this issue. Assuming the Fourier transform $\mathcal{F}\{\ldots\}$ yields a complex frequency spectrum:
\begin{equation}
    B(\nu)=\mathcal{F}\{ S_\mathrm{corr} \} \,,
\end{equation}
we can ensure the spectrum is real (or equivalently that the IGM is symmetric) by compensating the spectral phase term $\varphi^{-}_{\nu}$ using data from regions with non-zero intensity, i.e. where $|B(\nu)|\gg 0$. The spectral phase for the correction is defined as
\begin{equation}
    \varphi^{-}_{\nu} = \mathrm{tan}^{-1} \bigg( \frac{ \mathfrak{Im}\{ B(\nu) \} }{ \mathfrak{Re}\{ B(\nu)\}} \bigg) \,,
\end{equation}
where $\mathfrak{Im}\{ \ldots \}$ stands for the imaginary part of a complex argument. Finally, we find the corrected ILS-free frequency spectrum from
\begin{equation}
    B(\nu)_\mathrm{corr}=\mathfrak{Re}\{ B(\nu)\}\,\mathrm{cos}(\varphi^{-}_{\nu}) + \mathfrak{Im}\{ B(\nu) \}\, \mathrm{sin}(\varphi^{-}_{\nu}) \,.
\end{equation}
Such-treated spectra were used for absorbance calculations.

\vspace{0.2cm}
\noindent\textbf{Experimental setup details}: Light from the chip-scale comb devices was first collimated with a black-diamond high-numerical-aperture anti-reflective (AR) coated lens, and next guided to the inteferometer through a free-space Faraday optical isolator to prevent dynamic optical feedback effects resulting from the moving mirror. To record FTS IGMs, we used a Bruker Vertex 80 FTIR spectrometer with an external thermoelectrically-cooled photodetector (PVI-4TE-3.4, VIGO), whose near-DC output was conditioned by a a low-noise current preamplifier (SR570, Stanford Research Systems) set to filter signals below 100~kHz. The internal apertures of the FTS system were set to 1~mm. At a mirror modulation frequency of 40~kHz, 25.3~mm of OPD were scanned in 1~s, which accounting for the return of the mirror to the start position, communication with the instruments, and data exchange overhead, yielded 10 IGMs with 3--3.15~cm OPD per minute. Injection current stepping was provided by a precision source meter (Keithley, 2420) connected to the external modulation input port of a low-noise laser driver (D2-105-500, Vescent Photonics), which was also responsible for device's temperature stabilization. The ICL comb device was a single-section Fabry-P\'erot device with a 3~$\upmu$m wide ridge waveguide (Thorlabs) run without any frequency stabilization. The QWDL device described elsewhere~\cite{sterczewski2022battery} was analogously housed and biased, except for using a different laser driver due to the insufficient modulation range of the D2-105-500. In the latter case, we used an LDX-3620, ILX Lightwave low-noise laser driver capable of being modulated by tens of mA. Simultaneously with the optical IGMs, using a microwave spectrum analyzer we measured and recorded the intermode beat note spectrum extracted electrically from the device through a bias-T. 
\vspace{0.2cm}

\noindent\textbf{Data processing}: A Lorentzian fit was performed to all microwave intermode beat note spectra for $f_\mathrm{r}$ retrieval, while $f_0$ estimation followed the previously described protocols. To obtain the absorbance spectra (base-10 logarithm), we calculated the difference between two ILS-free spectra: with and without the analyte (see Supplementary Figures~S7 and S8 for raw sub-nominal resolution optical spectra). The absorbance spectra were next de-fringed using a sum-of-sines model due to parasitic etalons produced by the absorption cell's windows. For the methane, and acetylene measurements, simple point-wise division yielded effective cancellation of residual intensity versus current modulation produced by external cavity effects, which is also visible in the repetition rate and tuning characteristics of both comb platforms (see Supplementary Information, Section 2 and 5). For the HCN, due to the pronounced feedback sensitivity of the QWDL, the noise cancellation by means of point-wise division was less effective.

\vspace{0.2cm}

\footnotesize

\bibliographystyle{naturemag}
\bibliography{main}

\section*{Acknowledgements}
\footnotesize
\setstretch{1.}
\noindent
This work was supported under National Aeronautics and Space Agency’s (NASA) PICASSO program (106822 / 811073.02.24.01.85), and Research and Technology Development Spontaneous Concept Fund. It was in part performed at the Jet Propulsion Laboratory (JPL), California Institute of Technology, under contract with the NASA. L.~A. Sterczewski’s research was supported by an appointment to the NASA Postdoctoral Program at JPL, administered by Universities Space Research Association under contract with NASA. L.~A. Sterczewski acknowledges funding from the European Union's Horizon 2020 research and innovation programme under the Marie Skłodowska-Curie grant agreement No 101027721. The authors would like to thank Dr. Kevin Lascola, and Dr. Feng Xie at Thorlabs Inc. for providing the ICL material, and Dr. Clifford Frez at JPL for providing the diode laser material used in this study. Dr. Jerry Meyer, and Dr. Igor Vurgaftman at NRL are acknowledged for fruitful discussions on ICL combs.

\section*{Author contributions}
\footnotesize
\setstretch{1.}
\noindent L.A.S. conceived the idea. L.A.S. carried out the optical and electrical measurements and analyzed the data. L.A.S and M.B. wrote the manuscript. M.B. coordinated the project.

\section*{Conflict of interest}
\footnotesize
\setstretch{1.}
\noindent The authors have filed a provisional patent application on this idea, U.S. provisional application no. 63/418,717.  No other conflict of interest is present.

\section*{Data availability statement}
\footnotesize
\setstretch{1.}
\noindent The data that support the plots within this paper and other findings of this study are available from the corresponding author upon reasonable request.

\section*{Supplementary information}
\noindent Supplementary information accompanies this paper.

\end{document}


\vspace{-2cm}
\maketitle

\section{Interferogram processing}
\begin{figure}[!hp]
	\centering
	\includegraphics[width=1\textwidth]{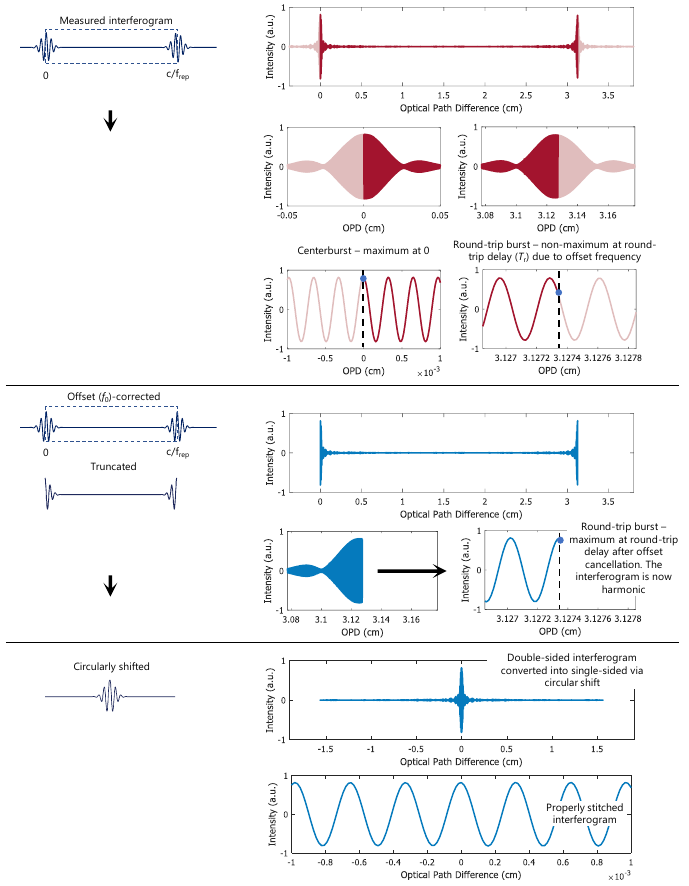}
	\caption{\textbf{Visual depiction of interferogram processing steps.}\label{fig:Schematic} }
\end{figure}

\vskip 0.2cm \noindent
Figure~\ref{fig:Schematic} visually depicts the evolution of the interferogram (IGM) at each processing step of the self-extracted sub-nominal resolution Fourier transform spectroscopy (FTS) routine. The left column shows a simulated IGM with a simplified temporal structure, while the right plots real data from a Fourier spectrometer. First, from a potentially longer IGM only the part between the peak of the centerburst ($\Delta=0$), and the satellite (round-trip) burst is kept (whose center position is retrieved from the measured repetition rate $f_\mathrm{r}$). The truncated part is plotted using a dark line. A non-zero offset frequency causes the round-trip burst to not have a maximum at the round-trip delay $T_\mathrm{r}=1/f_\mathrm{r}$. When the offset frequency is digitally canceled (through phase shifting of a Hilbert-transformed real IGM using a linear phase ramp), the IGM becomes harmonic with a local maximum at $T_\mathrm{r}$ (blue trace).

Such a trace is equivalent to an offset-free double-sided IGM acquired symmetrically around the zero-path-difference point (ZPD, $\Delta=0$). Its circular shift by $N_\mathrm{ss}$ samples (assuming the input is $N=2N_\mathrm{ss}+1$ samples long) yields the trace from the last row of Fig.~\ref{fig:Schematic}. Note that the two IGM sides match each other without any glitches or rapid jumps. This representation, however, is used only for visualization purposes. To calculate the spectrum, the non-center representation (asymmetric) is preferred.

\section{Tuning of the repetition rate and offset frequency -- diode combs at 3~$\upmu$m \label{sec:tuningf0frep}}
\vskip 0.2cm \noindent
The ability to precisely extract the offset frequency ($f_\mathrm{0}$) from the measured IGM coupled with a precise measurement of $f_\mathrm{r}$ using a microwave spectrum analyzer or frequency counter enables straightforward optical frequency axis retrieval. The high stability of the reference wavelength laser, which is used for measuring the optical displacement (or a precise optical encoder), can be directly leveraged for frequency axis calibration. This is particularly useful in more exotic spectral regions, where the availability of suitable single-mode lasers for anchoring the frequency axis is scarce. 

\begin{figure}[!ht]
	\centering
	\includegraphics[width=0.5\textwidth]{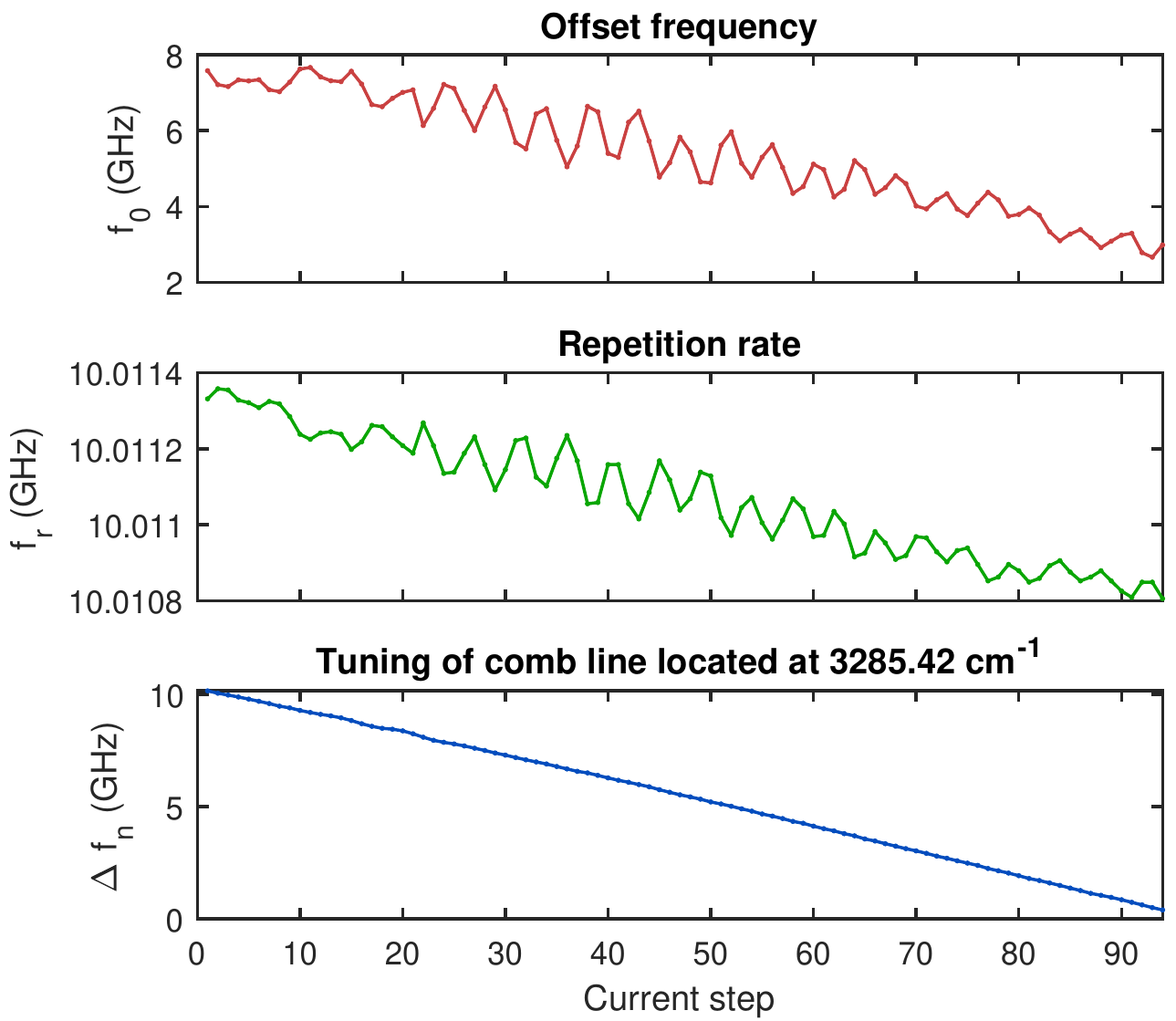}
	\caption{\textbf{Tuning of the offset frequency $f_0$, repetition rate $f_\mathrm{rep}$, and one of the comb lines for a quantum well diode laser frequency comb retrieved using the algorithm}. Note the anti-correlation between $f_0$ and $f_\mathrm{rep}$. Mutual cancellation of the oscillations yields nearly linear comb line position tuning.\label{fig:DiodeTuning} }
\end{figure}

Using the offset frequency retrieval routine described in detail in the main manuscript, we have characterized the offset frequency tuning behavior for two semiconductor laser frequency combs operating in the mid-infrared. Figure~\ref{fig:DiodeTuning} shows the anti-correlation of the tuning direction between $f_\mathrm{r}$ and $f_\mathrm{0}$ for a 3~$\upmu$m diode laser frequency comb. Whereas $f_\mathrm{r}$ is \emph{measured} with a microwave spectrum analyzer, $f_\mathrm{0}$ is \emph{estimated}. The oscillatory shape of the two curves results most likely from external optical cavity effects. During the injection current scan, when comb lines were tuned by a full $f_\mathrm{r}$, both frequencies suffered from periodic oscillations in counter phase. A coarse estimate of 20 oscillation periods over $f_\mathrm{r}\approx10$~GHz yields a single oscillation period of $\sim$500~MHz. This corresponds to an external cavity round trip length of 60~cm, which can be attributed to the position of the optical isolator (sub-optimal for this wavelength), located $\sim$30~cm away from the laser facet.

\section{Offset frequency ambiguity}
\begin{figure}[!ht]
	\centering
	\includegraphics[width=0.5\textwidth]{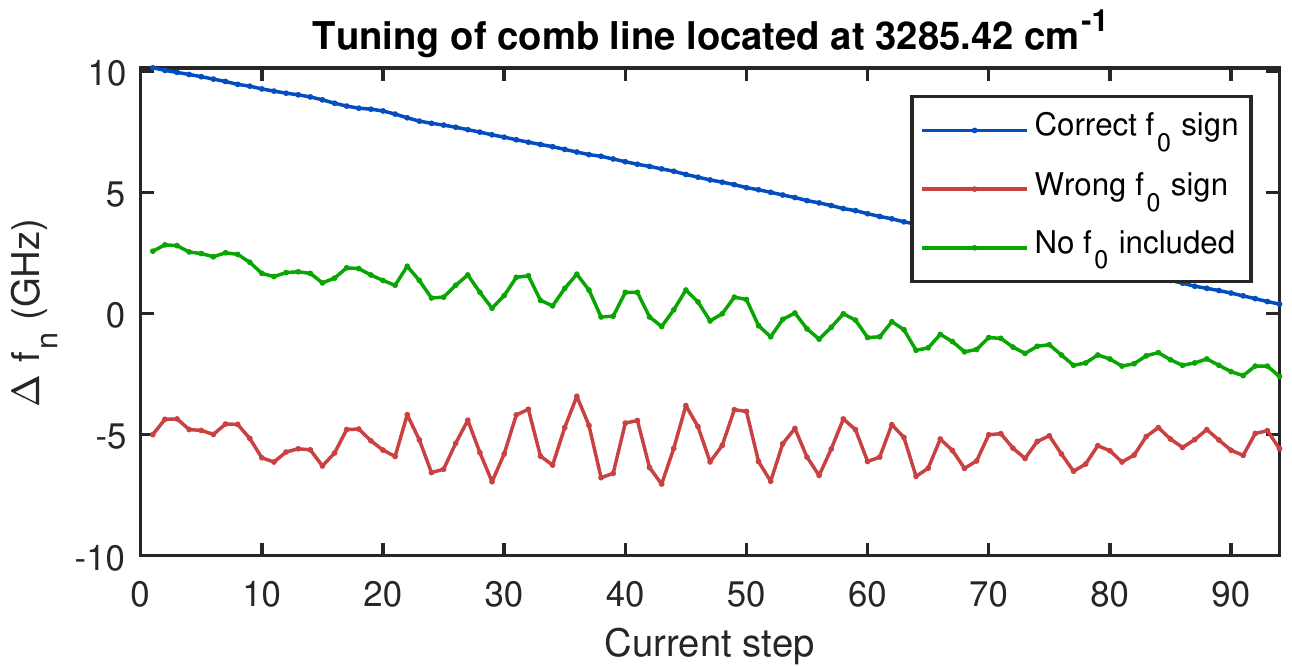}
	\caption{\textbf{The problem of $f_0$ ambiguity can be easily solved for scanned measurements.} Prior knowledge about the nearly linear tuning of the optical wavelength/frequency can be incorporated to determine the sign of $f_0$.\label{fig:DiodeTuningWrong} }
\end{figure}

\vskip 0.2cm \noindent
It is imperative to discuss the offset frequency tuning characteristics in the context of sign ambiguity ($\pm f_0$). For semiconductor lasers, one expects 
nearly linear optical frequency tuning with injection current. Figure~\ref{fig:DiodeTuningWrong} shows that such behavior is obtained only in one case -- when the offset frequency decreases with injection current (here corresponding to $+f_0$). In the opposite case ($-f_0$), or when it is not included in the frequency retrieval formula, incorrect and highly oscillatory comb line tuning characteristics are obtained. Experimental proof for line tuning linearity obtained with an independent optical instrument is provided in Section~\ref{sec:diodeTuningLinearity}.

\newpage

\section{Tuning of optical spectra characterized using an optical spectrum analyzer \label{sec:diodeTuningLinearity}}
\begin{figure}[!h]
\centering
\begin{subfigure}{.37\textwidth}
\centering
  \includegraphics[height=5cm]{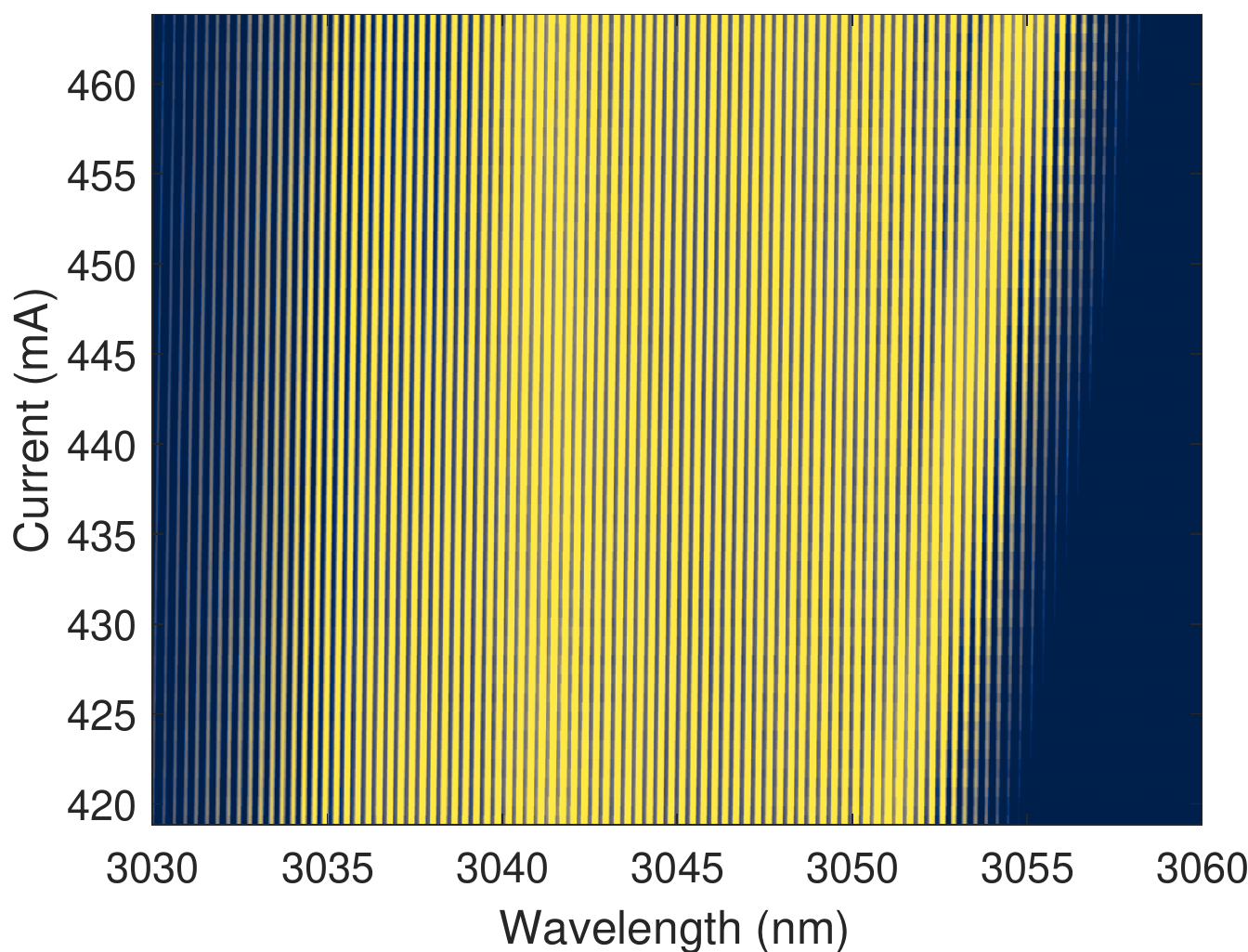}  
  \caption{Full-span optical spectrum}
  \label{fig:diodeTuningUpA}
\end{subfigure}
\begin{subfigure}{.37\textwidth}
  \centering
  \includegraphics[height=5cm]{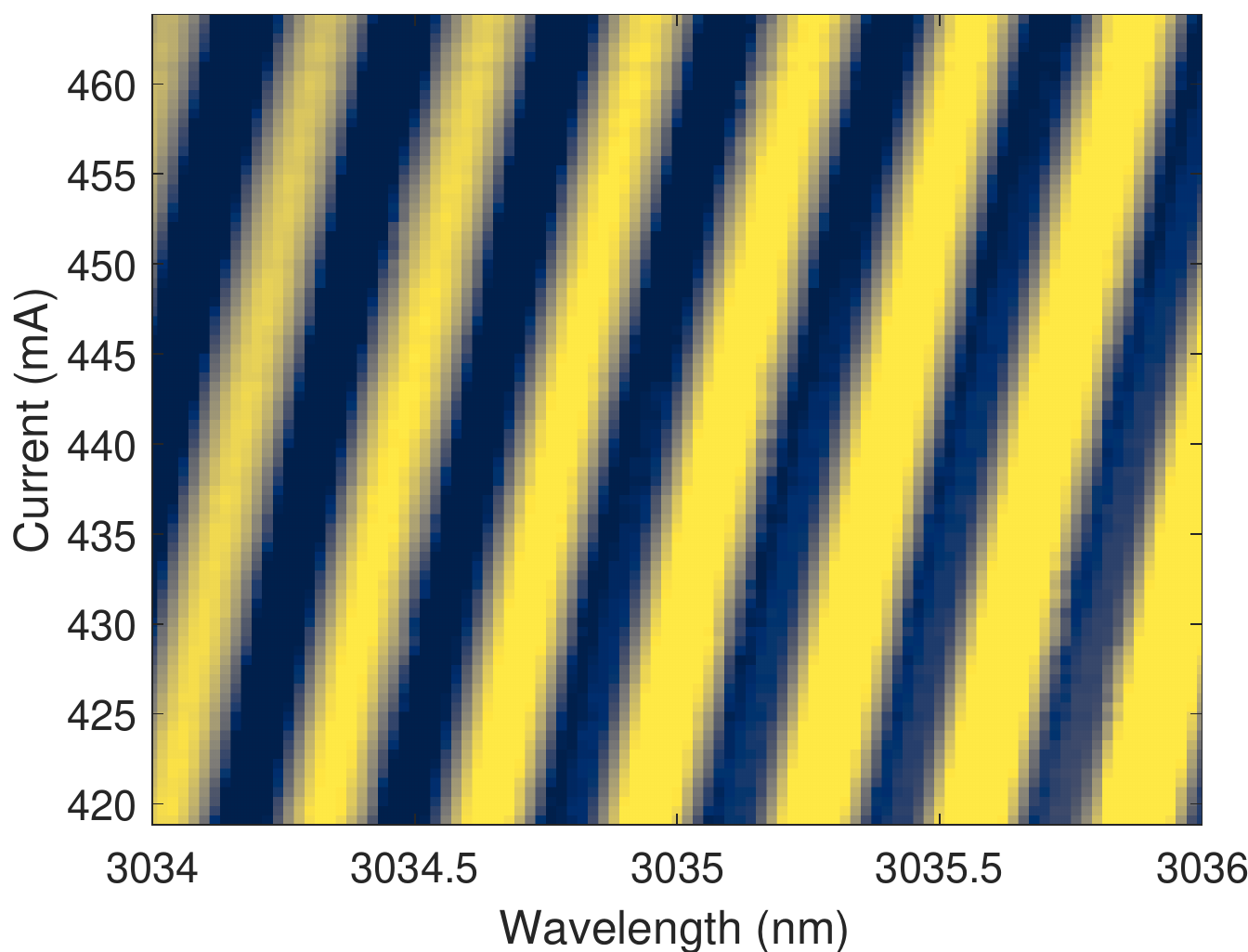}  
  \caption{Zoomed shorter-wavelength part}
  \label{fig:diodeTuningUpB}
\end{subfigure}
\begin{subfigure}{.18\textwidth}
\centering
  \includegraphics[height=5cm]{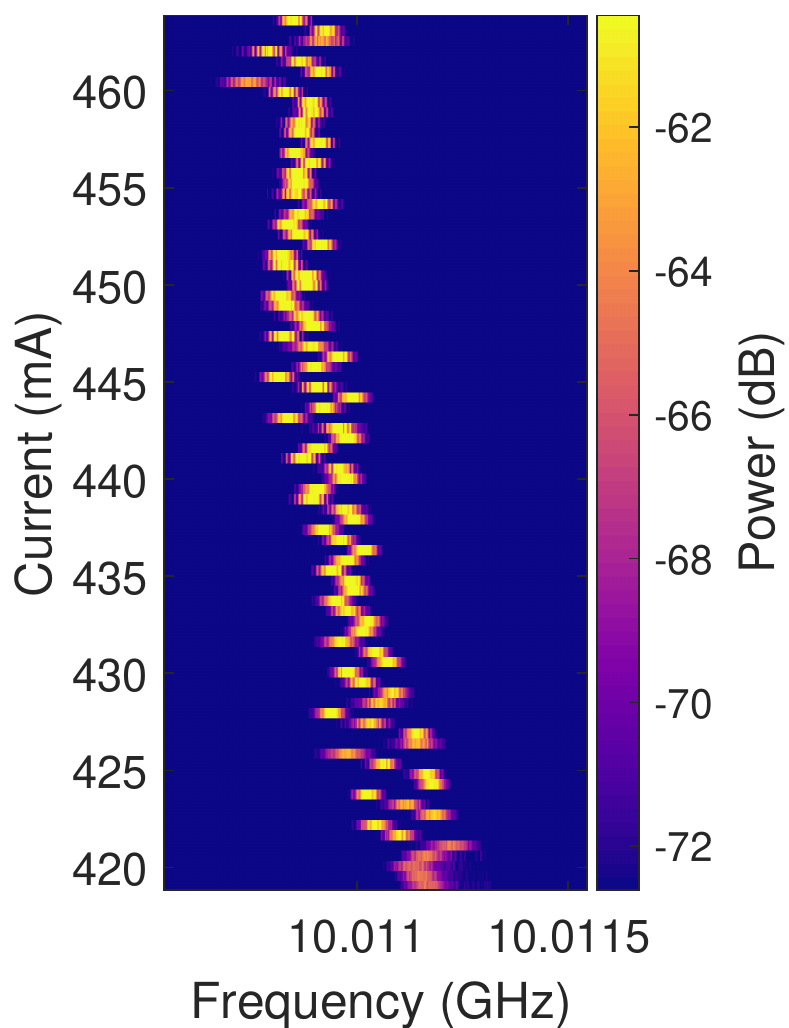}  
  \caption{Radio-frequency spectra}
  \label{fig:diodeTuningUpC}
\end{subfigure}
\caption{ \textbf{Tuning capabilities of the 3$~\upmu$m wavelength diode comb for gap-less high-resolution spectroscopy.} The spectra were measured for diagnostic purposes with an optical spectrum analyzer (a--b) and a microwave spectrum analyzer (c). The wavelength increases almost linearly with injection current, which corresponds to a decrease of the optical frequency (wavenumber). A full free spectral range (FSR) scan required increasing the injection current by 44.5~mA ($\sim$10\%). Note the oscillatory tuning of the RF beat note. \label{fig:diodeTuningUp}}
\end{figure}

\begin{figure}[!h]
\centering
\begin{subfigure}{.37\textwidth}
\centering
  \includegraphics[height=5cm]{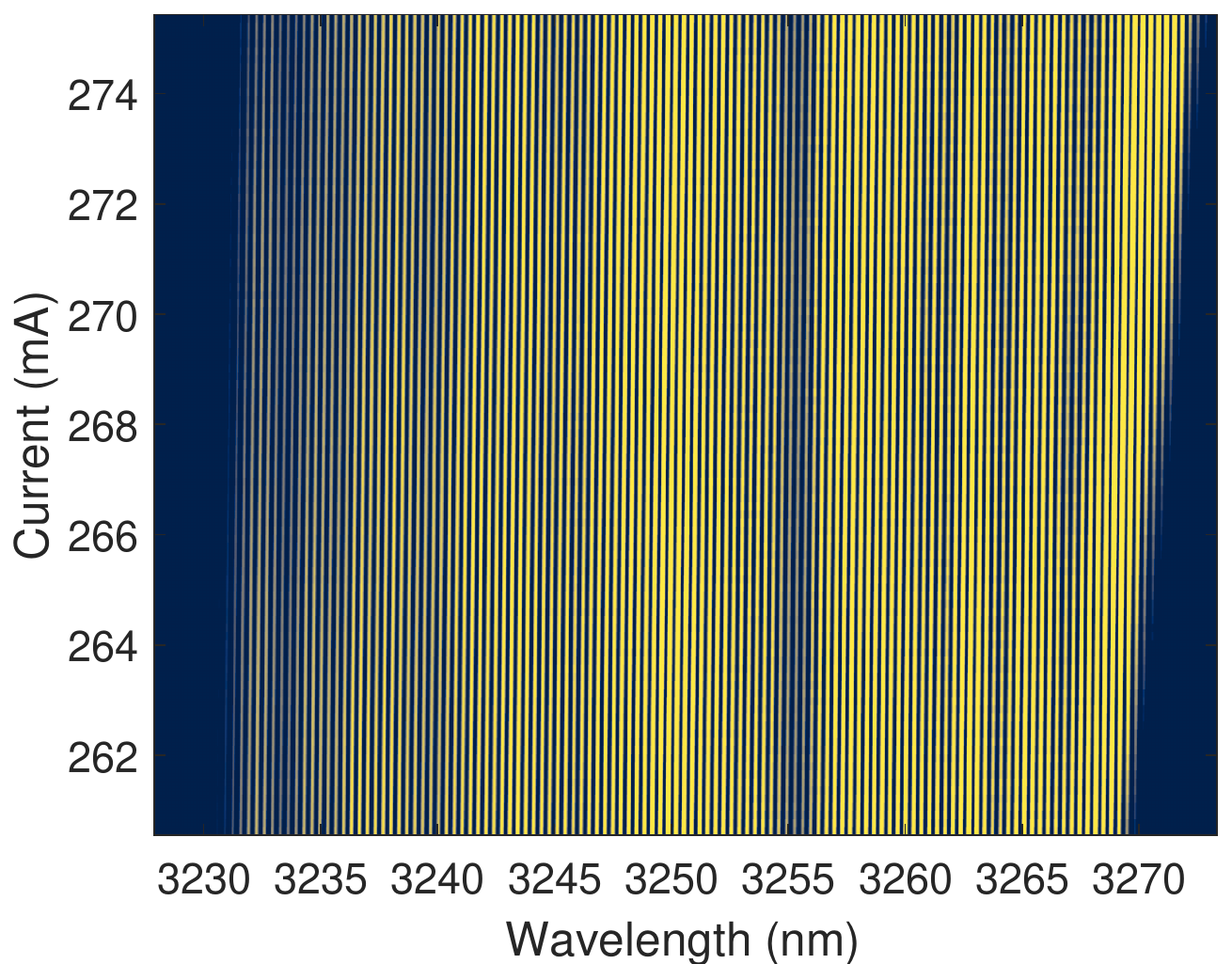}  
  \caption{Full-span optical spectrum}
  \label{fig:iclTuningUpA}
\end{subfigure}
\begin{subfigure}{.37\textwidth}
  \centering
  \includegraphics[height=5cm]{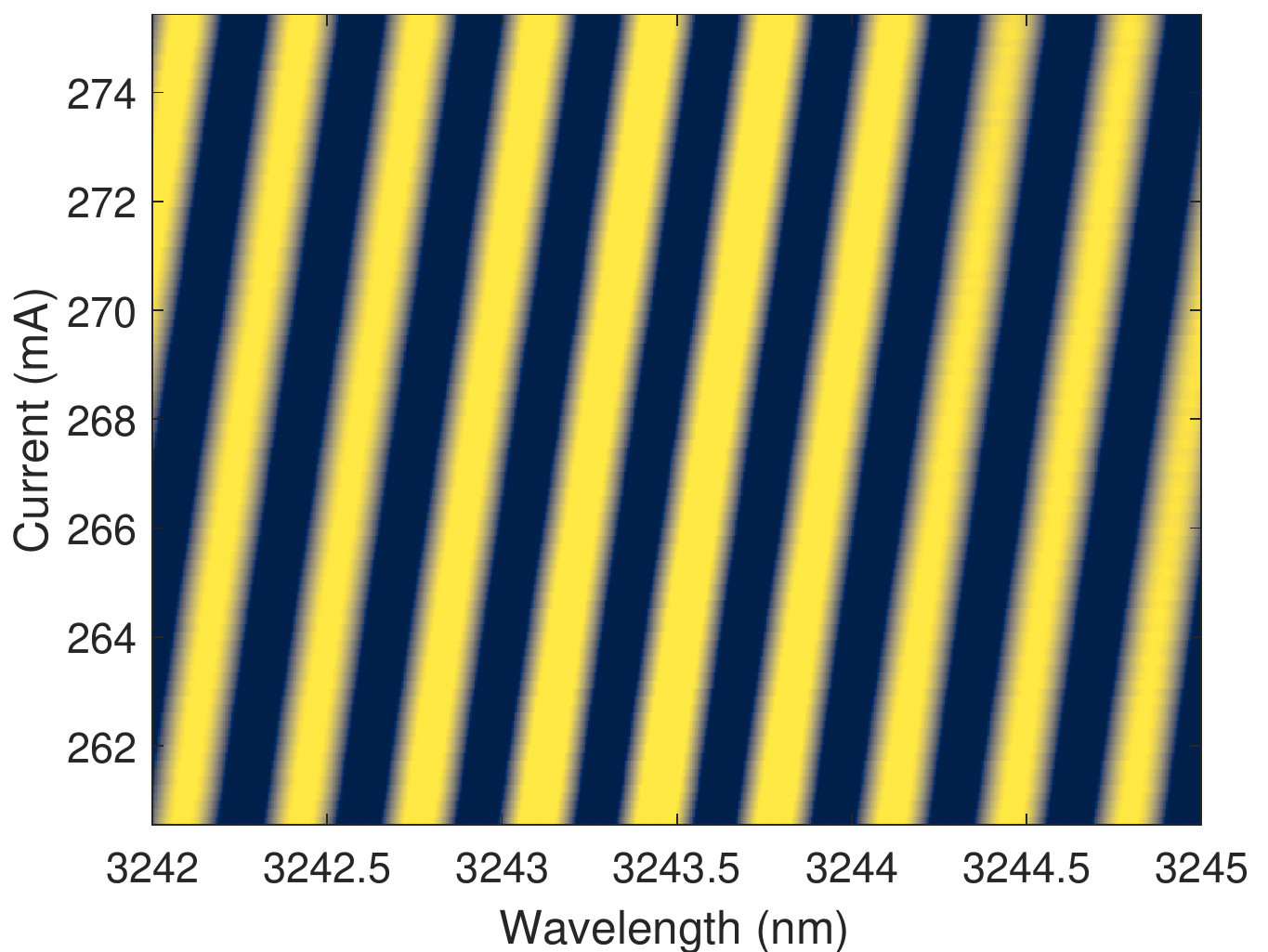}  
  \caption{Zoomed shorter-wavelength part}
  \label{fig:iclTuningUpB}
\end{subfigure}
\begin{subfigure}{.18\textwidth}
\centering
  \includegraphics[height=5cm]{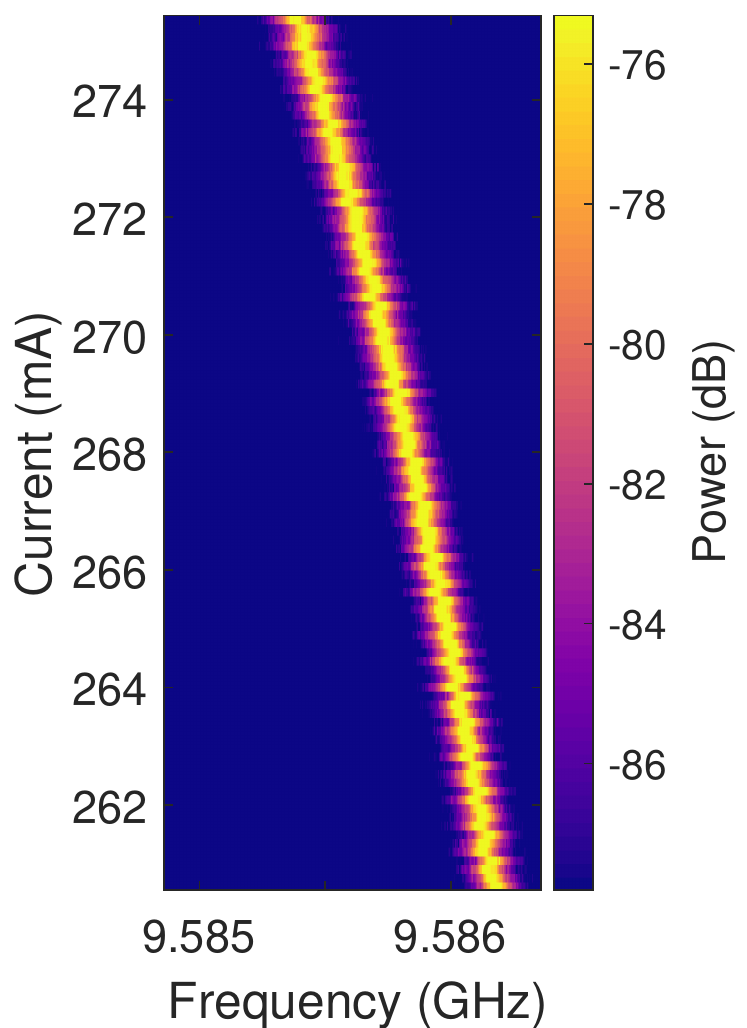}  
  \caption{Radio-frequency spectra}
  \label{fig:iclTuningUpC}
\end{subfigure}
\caption{ \textbf{Tuning capabilities of the 3.25$~\upmu$m wavelength interband cascade laser frequency comb for gap-less high-resolution spectroscopy.} The spectra were measured for diagnostic purposes with a Fourier spectrometer (a--b) and a microwave spectrum analyzer (c). The wavelength increases almost linearly with the injection current, which corresponds to a decrease in the optical frequency (wavenumber). A full free spectral range (FSR) scan required increasing the injection current by 14.7~mA ($\sim$5\%). \label{fig:iclTuningUp}}
\end{figure}

\vskip 0.2cm \noindent
Experimental evidence of comb line tuning linearity despite the oscillatory trajectories of $f_\mathrm{r}$ is shown in Fig.~\ref{fig:diodeTuningUp} and Fig.~\ref{fig:iclTuningUp}. Two comb platforms are characterized here. Figure~\ref{fig:diodeTuningUp} shows the tuning map for diode laser frequency combs, while Fig.~\ref{fig:iclTuningUp} shows one for an interband cascade laser frequency comb. It is clear that the diode comb is exhibits richer $f_\mathrm{r}$ tuning dynamics. 

\newpage

\section{Tuning of the repetition rate and offset frequency - ICLs at 3.25~$\upmu$m}
\begin{figure}[!ht]
	\centering
	\includegraphics[width=0.5\textwidth]{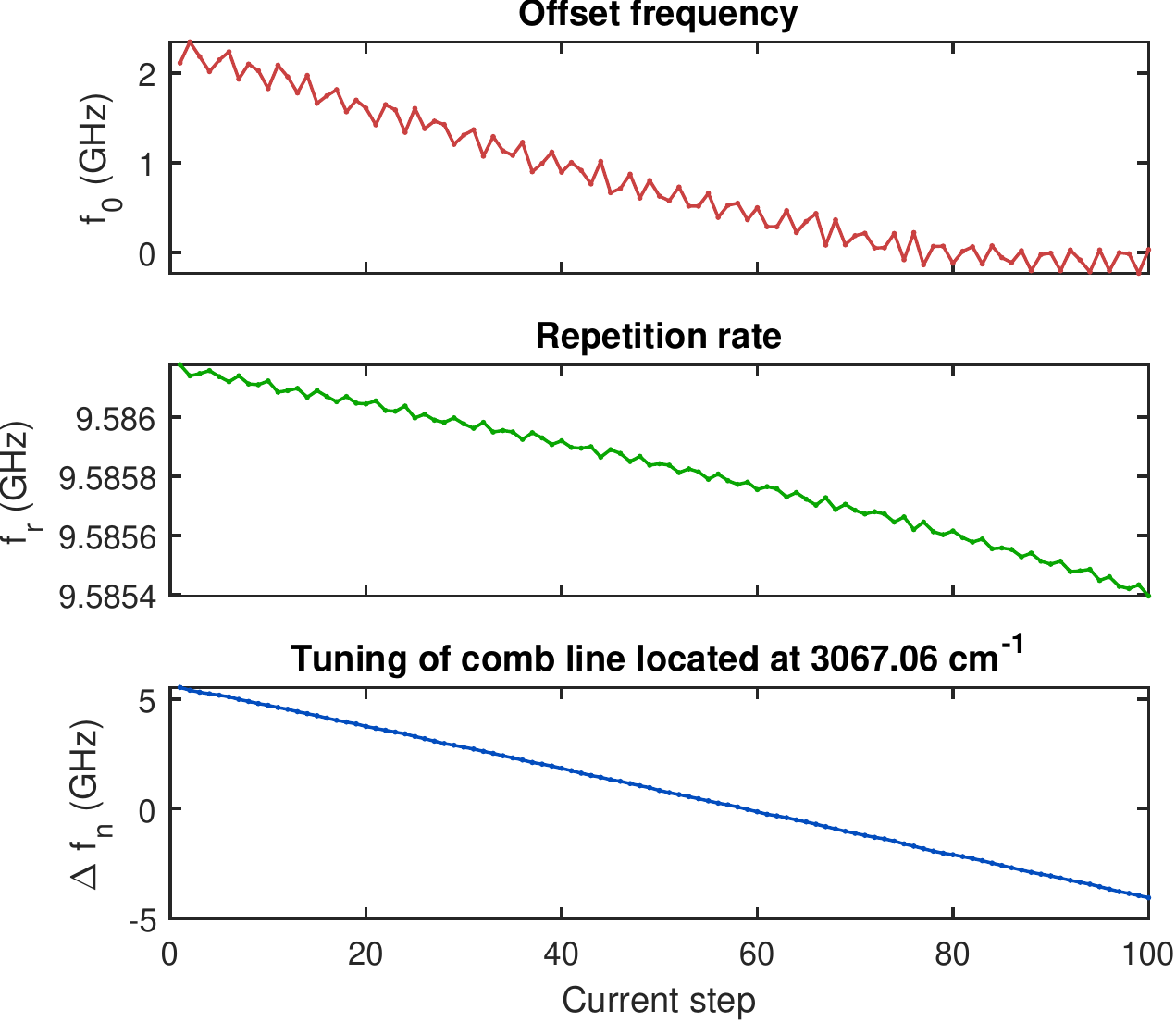}
	\caption{\textbf{Optical frequency axis retrieval for ICL combs}. Tuning of the offset frequency $f_0$, repetition rate $f_\mathrm{rep}$, and one of the comb lines for an interband cascade laser frequency comb retrieved using the algorithm. Note the anti-correlation between $f_0$ and $f_\mathrm{rep}$. Mutual cancellation of the oscillations yields nearly linear comb line position tuning. \label{fig:ICLTuning} }
\end{figure}

\vskip 0.2cm \noindent
Analogously to the data presented in Section~\ref{sec:tuningf0frep}, we have retrieved the frequency tuning characteristics for ICL combs operating with a center wavelength of 3.25~$\upmu$m. This is shown in Fig.~\ref{fig:ICLTuning}. Note the much lower amplitude of frequency oscillations of $f_0$ and $f_\mathrm{r}$ because the emission wavelengths almost match the optimal wavelength of the coating and one that provides the best optical isolation from downstream optical components (3.35~$\upmu$m). However, the periodicity of the oscillations is halved compared to the diode laser comb case ($\sim$36 periods yielding a period of$\sim$267~MHz). This corresponds to an external cavity roundtrip length of $\sim$1.12~m, which can be attributed to the position of another parasitic etalon in the system -- the windows of the gas cell.

\newpage

\section{Raw subnominal-resolution optical spectra}
\begin{figure}[!h]
\centering
\begin{subfigure}{.55\textwidth}
  \centering
  \includegraphics[width=\textwidth]{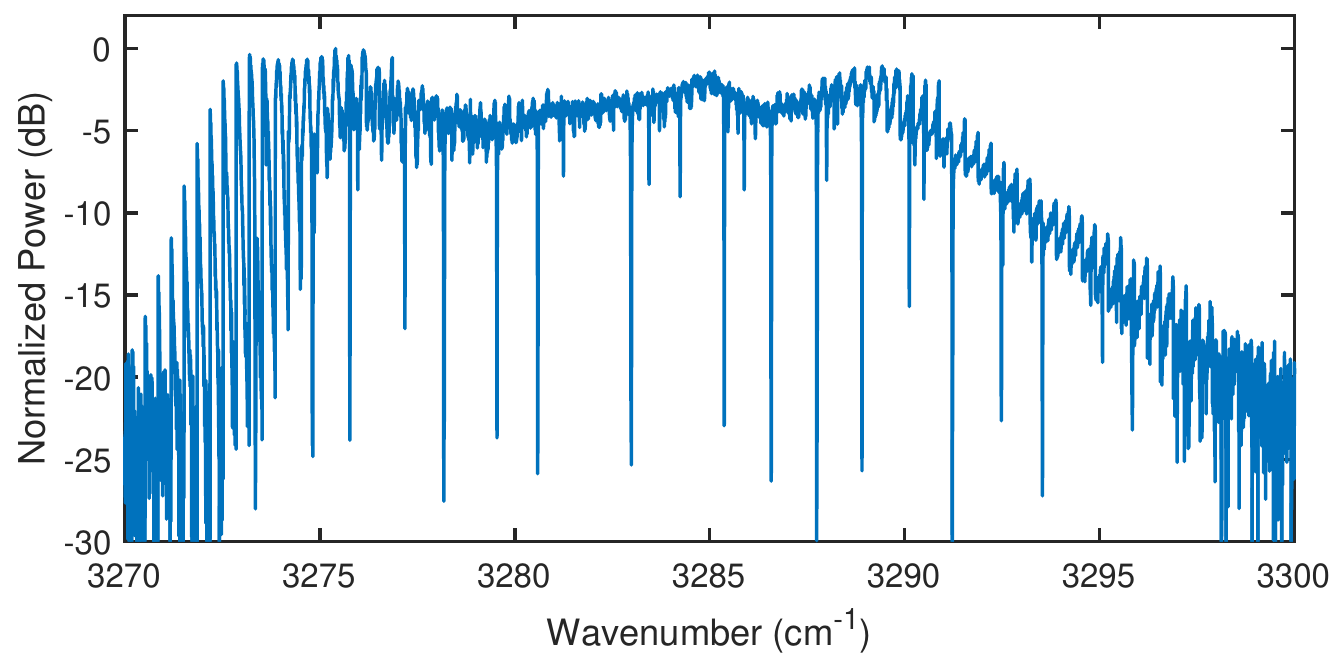}  
  \caption{}
  \label{fig:combinedSpectrumA}
\end{subfigure}

\begin{subfigure}{.55\linewidth}
\centering
  \includegraphics[width=\textwidth]{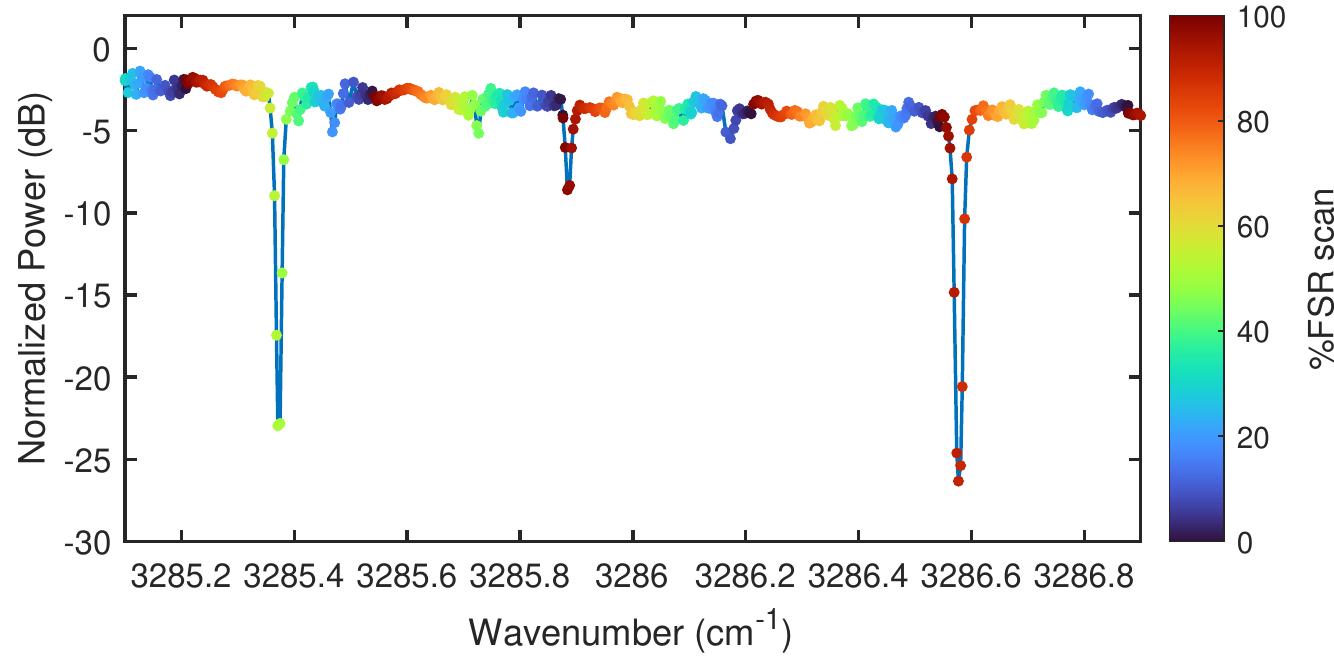}  
  \caption{}
  \label{fig:combinedSpectrumB}
\end{subfigure}

\begin{subfigure}{.55\textwidth}
\centering
  \includegraphics[width=\textwidth]{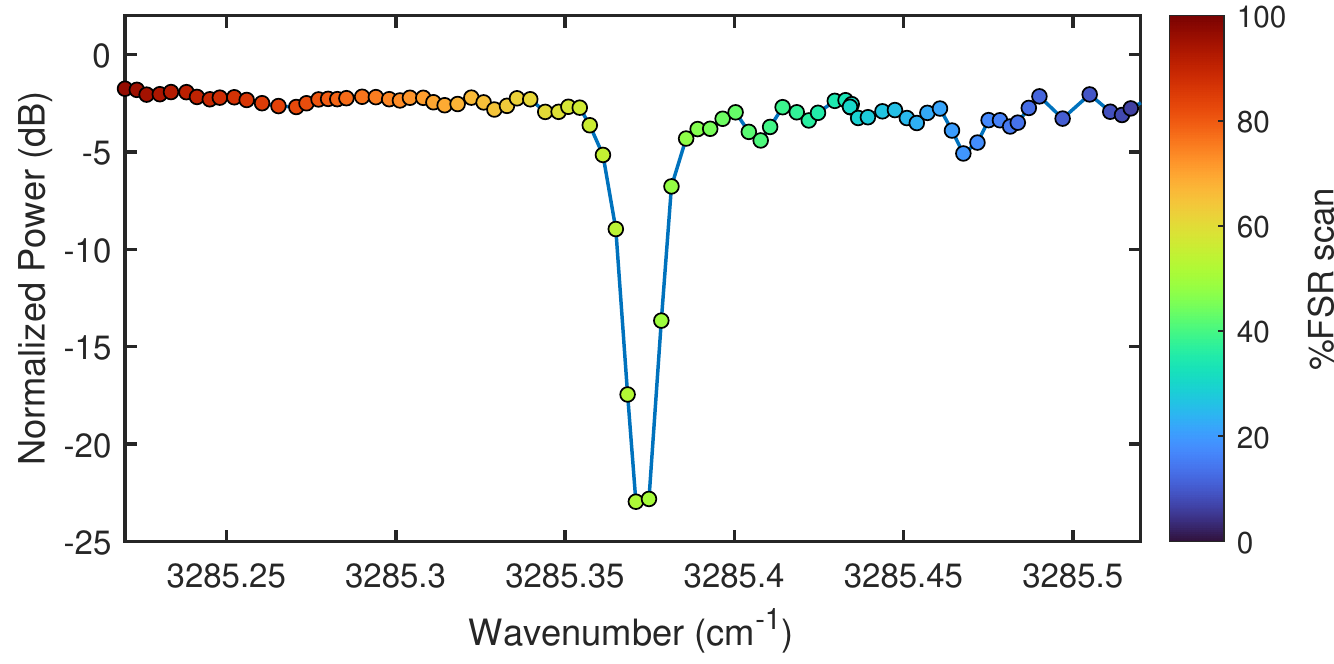}  
  \caption{}
  \label{fig:combinedSpectrumC}
\end{subfigure}

\begin{subfigure}{.55\textwidth}
\centering
  \includegraphics[width=\textwidth]{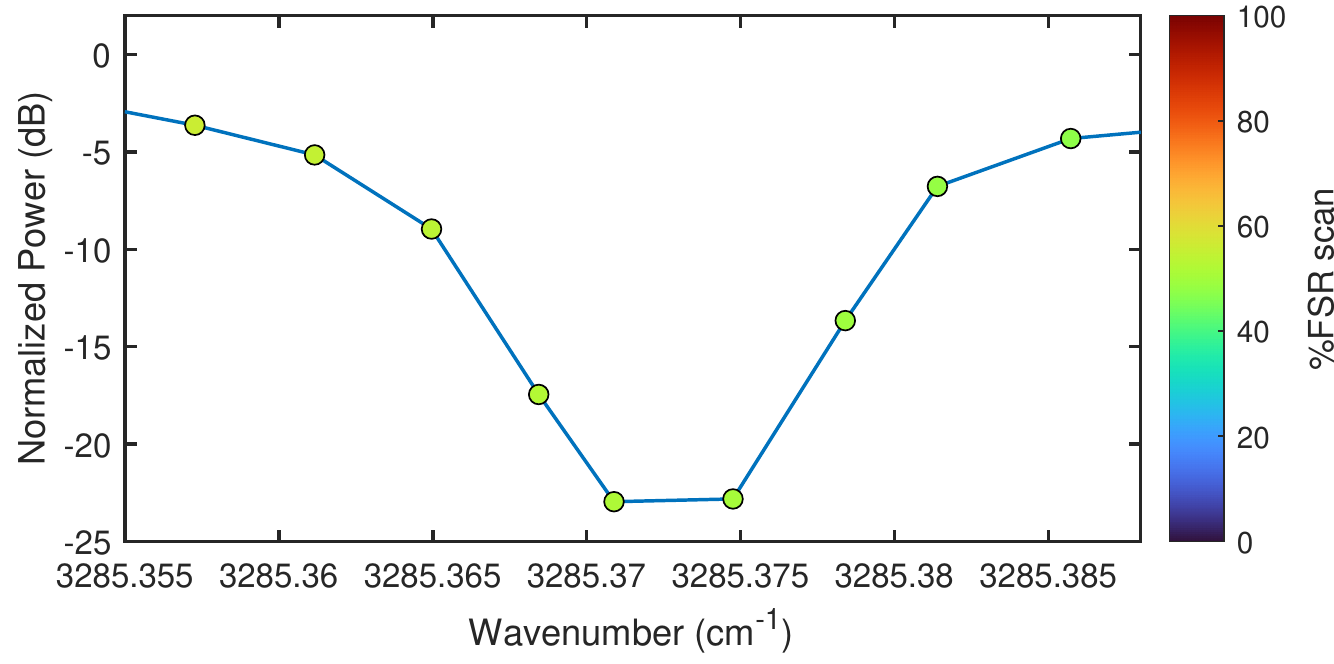}  
  \caption{}
  \label{fig:combinedSpectrumD}
\end{subfigure}
\caption{ \textbf{Sub-nominal resolution FTS spectra showing the tuning range of individual comb lines. The analyte was low pressure acetylene (C$_2$H$_2$ at 2 Torr).} The sawtooth-like spectral shape on the edges of the spectrum results from its progressive broadening and appearance of comb lines that do not exist at lower injection currents. (b) Span of 1.8~cm$^{-1}$ (54~GHz), (c) Span of 0.3~cm$^{-1}$ (9~GHz), (d) Span of 0.033~cm$^{-1}$ (1000~MHz).\label{fig:combinedSpectrumC2H2}}
\end{figure}

\begin{figure}[!h]
\centering
\begin{subfigure}{.55\textwidth}
  \centering
  \includegraphics[width=\textwidth]{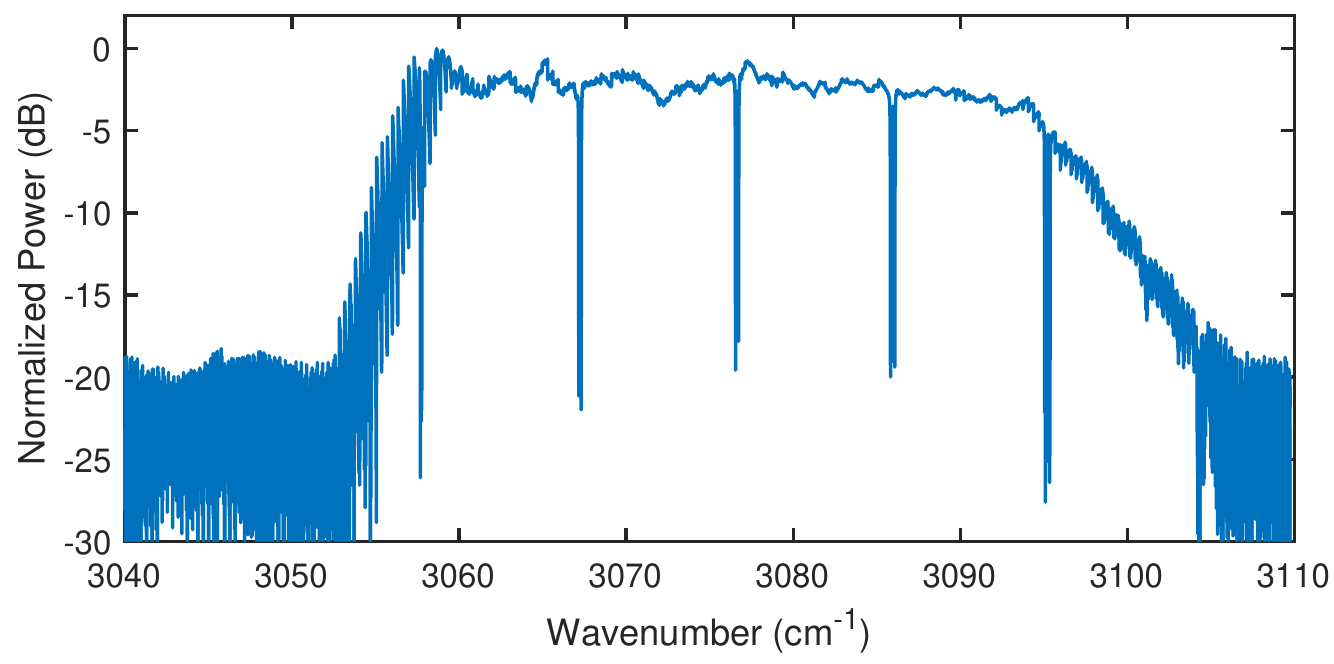}  
  \caption{}
  \label{fig:CH4combinedSpectrumA}
\end{subfigure}

\begin{subfigure}{.55\linewidth}
\centering
  \includegraphics[width=\textwidth]{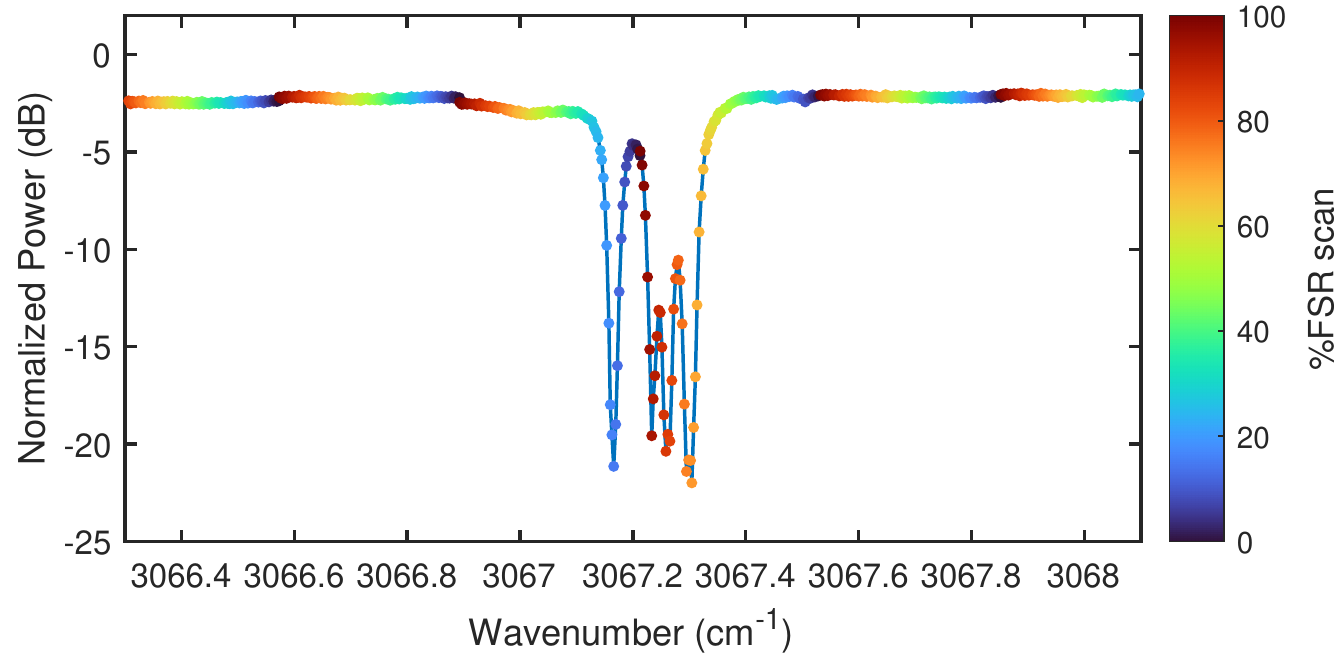}  
  \caption{}
  \label{fig:CH4combinedSpectrumB}
\end{subfigure}

\begin{subfigure}{.55\textwidth}
\centering
  \includegraphics[width=\textwidth]{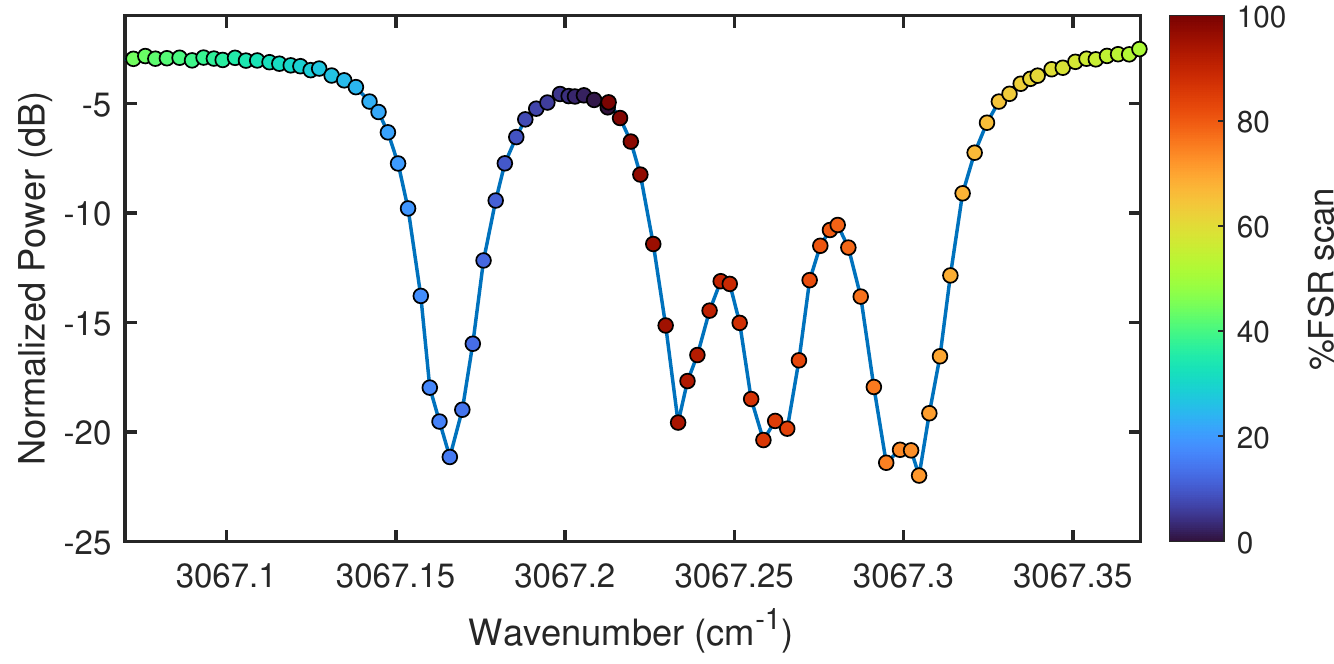}  
  \caption{}
  \label{fig:CH4combinedSpectrumC}
\end{subfigure}

\begin{subfigure}{.55\textwidth}
\centering
  \includegraphics[width=\textwidth]{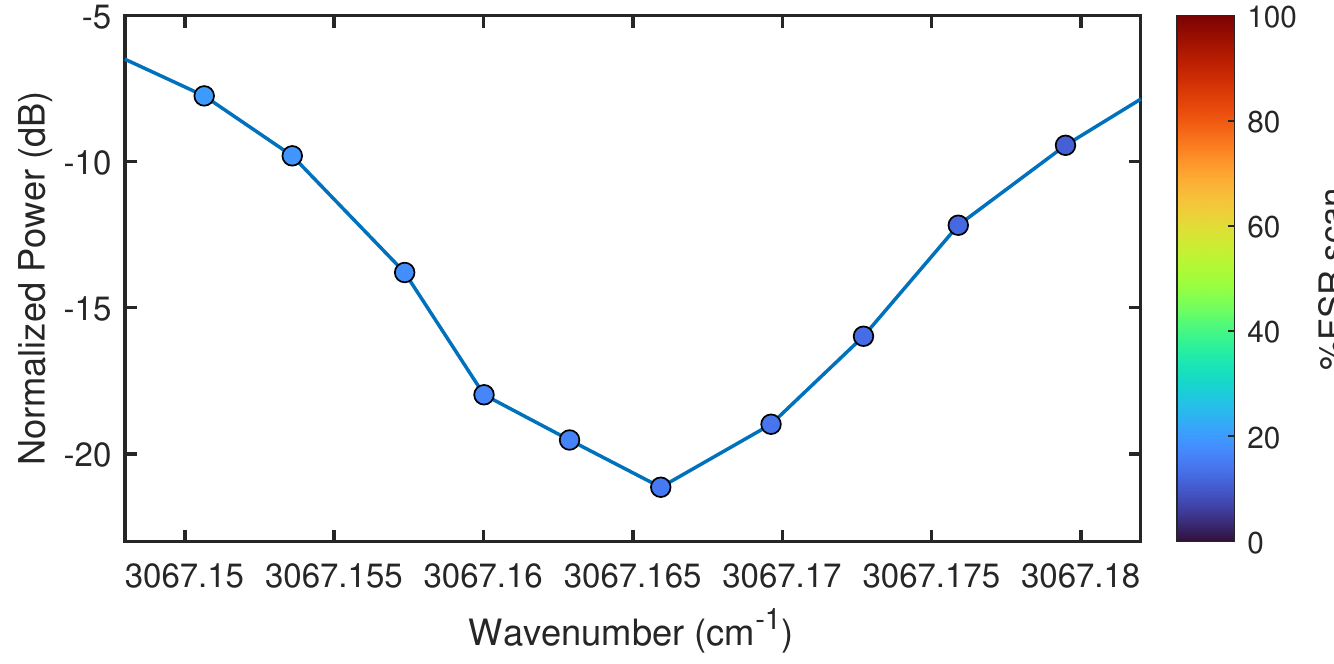}  
  \caption{}
  \label{fig:CH4combinedSpectrumD}
\end{subfigure}
\caption{ \textbf{Sub-nominal resolution FTS spectra showing the tuning range of individual comb lines. The analyte was methane.} The sawtooth-like spectral shape on the edges of the spectrum results from its progressive broadening and appearance of comb lines that do not exist at lower injection currents. Panels (b--d) show individual modal intensities during the current scan. Dots sharing the same color are spaced by $f_\mathrm{r}$. (b) Span of 1.8~cm$^{-1}$ (54~GHz), (c) Span of 0.3~cm$^{-1}$ (9~GHz), (d) Span of 0.034~cm$^{-1}$ (1020~MHz). \label{fig:combinedSpectrumCH4}}
\end{figure}

\vskip 0.2cm \noindent
For reader's convenience, and justification of the $\sim$20~dB dynamic range, Fig.~\ref{fig:combinedSpectrumC2H2} and Fig.~\ref{fig:combinedSpectrumCH4} plot raw interleaved spectra calculated using the subnominal resolution routine for acetylene (C$_2$H$_2$), and methane (CH$_4$), respectively. Panels (a) show full-span interleaved spectra with frequency axis calibration based on the estimated $f_0$, known $\lambda_\mathrm{ref}$ (temperature stabilized HeNe laser) and measured $f_\mathrm{r}$. Panels (b) are scatter-type plots where dots sharing the same color correspond to individual comb lines coexisting at a given injection current. They are separated by 0.32--0.33~cm$^{-1}$, which is the comb repetition rate expressed in wavenumbers. Injection current tuning responsible for spectral interleaving is analogous to having $\sim$100 single-mode lasers simultaneously tuned in frequency. Each tuning curve is uniquely defined by the comb line number $n$, $f_0$, and $f_\mathrm{r}$. Note that the saw-tooth-like shape of the spectral edges is caused by the appearance of new comb lines that do not exist at lower injection currents. In other words, with increased pumping, the spectrum gradually broadens and lines of the spectral edges reach higher intensities. In contrast, the central part of the spectrum is dominated by frequency tuning with minor intensity changes.

